\documentclass[%
superscriptaddress,
anofootinbib,
 amsmath,amssymb,
 aps,
]{revtex4-2}

\usepackage{graphicx}% Include figure files
\usepackage{dcolumn}% Align table columns on decimal point
\usepackage{bm}% bold math
\usepackage{hyperref}% add hypertext capabilities
\usepackage{url}
\begin{document}

\title{Electron Collision Cross Sections in Tetrafluoropropene HFO1234ze(E) for Gas Mixtures in Resistive Plate Chambers}

%% use optional labels to link authors explicitly to addresses:
%% \author[label1,label2]{}
%% \address[label1]{}
%% \address[label2]{}

\author{Antonio Bianchi} \email{antonio.bianchi@mi.infn.it}
%\altaffiliation[Also at ]{Physics Department, XYZ University.}%Lines break automatically or can be forced with \\
\affiliation{INFN, Milan, Italy}

\author{Alessandro Ferretti}
\affiliation{Università degli Studi di Torino and INFN, Turin, Italy}

\author{Martino Gagliardi}
\affiliation{Università degli Studi di Torino and INFN, Turin, Italy}

\author{Ermanno Vercellin}
\affiliation{Università degli Studi di Torino and INFN, Turin, Italy}

\begin{abstract}
%% Text of abstract
In recent years, there has been growing interest in tetrafluoropropene HFO1234ze(E) (C$_{3}$H$_{2}$F$_{4}$) for Resistive Plate Chambers (RPCs). This novel gas is considered a promising alternative to the standard mixtures currently used in RPCs, thanks to its low global warming potential. The knowledge of electron collision cross sections in C$_{3}$H$_{2}$F$_{4}$ enables reliable predictions of electron transport coefficients and reaction rates in C$_{3}$H$_{2}$F$_{4}$-based gas mixtures. This allows for optimizing the C$_{3}$H$_{2}$F$_{4}$-based gas mixtures to achieve the desired performance in RPCs.

From measurements of electron transport coefficients and reaction rates, a complete set of scattering cross sections for electrons in C$_{3}$H$_{2}$F$_{4}$ has been derived. Validation of the electron collision cross sections is achieved through systematic comparisons of electron swarm parameters with experimental data in both pure C$_{3}$H$_{2}$F$_{4}$ and C$_{3}$H$_{2}$F$_{4}$/CO$_{2}$ gas mixtures. Given the influence of electron attachment in C$_{3}$H$_{2}$F$_{4}$ by the gas density, this work also includes precise calculations of the critical electric field strength in such mixtures. This set of cross sections has been further utilized to compute the effective ionization Townsend coefficient in gas mixtures containing C$_{3}$H$_{2}$F$_{4}$, potentially applicable for RPCs. 

%%The aim is to predict the values of the reduced electric field at which electron avalanches start to develop in RPCs when operated with C$_{3}$H$_{2}$F$_{4}$-based gas mixtures.

\end{abstract}

%%Graphical abstract
%%\begin{graphicalabstract}
%\includegraphics{grabs}
%%\end{graphicalabstract}

%%Research highlights
%%\begin{highlights}
%%\item Research highlight 1
%%\item Research highlight 2
%%\end{highlights}

%\keywords{Suggested keywords}%Use showkeys class option if keyword
                              %display desired
\maketitle

%\tableofcontents

\section{Introduction}
Tetrafluoropropene HFO1234ze(E) (C$_{3}$H$_{2}$F$_{4}$) is a cooling gas that has been introduced as a potential alternative of tetrafluoroethane C$_{2}$H$_{2}$F$_{4}$ in a number of refrigerant applications, thanks to its good cooling capabilities and low global warming potential (GWP). Unlike C$_{2}$H$_{2}$F$_{4}$, whose GWP is more than 1300 \cite{IPCC}, the tetrafluoropropene C$_{3}$H$_{2}$F$_{4}$ has a carbon-carbon double bound in its chemical structure that makes it more reactive in the atmosphere than C$_{2}$H$_{2}$F$_{4}$. This implies the reduction of its atmospheric lifetime in comparison to that of C$_{2}$H$_{2}$F$_{4}$ and, as a consequence, a lower GWP. According to the European Regulation \cite{EUregulation} for restrictions on the usage of fluorinated gases, the GWP of C$_{3}$H$_{2}$F$_{4}$ is evaluated to be 7 \cite{EUregulation}, while latest studies have revised this value to less than 1 \cite{IPCC}.

In recent years, several studies have investigated the possibility of substituting C$_{2}$H$_{2}$F$_{4}$ with C$_{3}$H$_{2}$F$_{4}$ in gaseous particle detectors, especially in Resistive Plate Chambers (RPCs) \cite{santonico1981development} that are operated with gas mixtures containing high concentrations of C$_{2}$H$_{2}$F$_{4}$. In particular, gas mixtures of C$_{3}$H$_{2}$F$_{4}$ and CO$_{2}$ exhibit significant potential in RPCs \cite{abbrescia2016preliminary, bianchi2019characterization, rigoletti2020studies, proto2022new, rigoletti2023studies}. The usage of C$_{3}$H$_{2}$F$_{4}$ has not been considered only in gaseous particle detectors, but also in voltage insulation applications \cite{koch2015high, alise2017proceedings} as a replacement of the much less environmental-friendly sulfur hexafluoride (SF$_{6}$).

The increasing interest in C$_{3}$H$_{2}$F$_{4}$ for gaseous particle detectors and voltage insulating applications led to the evaluation of electron transport coefficients and reaction rates in this gas. Indeed, electron swarm parameters in pure C$_{3}$H$_{2}$F$_{4}$ and C$_{3}$H$_{2}$F$_{4}$-based gas mixtures have been recently measured by Chachereau et al. \cite{alise2017proceedings, alise2016paper} thanks to a pulsed Townsend experiment \cite{dahl2012obtaining}, whereas no electron collision cross sections in C$_{3}$H$_{2}$F$_{4}$ are available in the literature so far.

This paper aims to present a set of scattering cross sections for electrons in C$_{3}$H$_{2}$F$_{4}$ that allows reliable predictions of electron transport coefficients and reaction rates in pure C$_{3}$H$_{2}$F$_{4}$ and mixtures of C$_{3}$H$_{2}$F$_{4}$/CO$_{2}$ in different concentrations. The knowledge of electron collision cross sections in C$_{3}$H$_{2}$F$_{4}$ enables predicting the efficiency curves of RPCs and optimizing gas mixtures containing C$_{3}$H$_{2}$F$_{4}$ to achieve the desired performance in RPCs. The set of electron collision cross sections has been derived thanks to the implementation of an iterative technique performed on the experimental data, available in the literature at present. The well-established technique has been already utilized for numerous gas molecules, including those containing multiple fluorine atoms, such as C$_{2}$H$_{2}$F$_{4}$ \cite{sasic2013scattering}, C$_{4}$F$_{7}$N \cite{zhang2023determination} and CF$_{3}$I \cite{kimura2010electron}.

The paper is organized as follows. Section \ref{sec:techniques} gives a brief overview of the numerical techniques to calculate the electron swarm parameters starting from a generic set of electron collision cross sections. The iterative procedure used to unfold the experimental swarm data, aiming to obtain the electron collision cross sections in C$_{3}$H$_{2}$F$_{4}$, is described in section \ref{sec:methodology}. Section \ref{sec:results} shows the results of this procedure. In particular, a systematic comparison between calculations and measurements of electron swarm parameters in pure C$_{3}$H$_{2}$F$_{4}$ at different values of pressure is outlined in section \ref{sec:results1} while additional comparisons of these parameters in gas mixtures of C$_{3}$H$_{2}$F$_{4}$/CO$_{2}$ and C$_{3}$H$_{2}$F$_{4}$/Ar are in sections \ref{sec:results2} and \ref{sec:results3}, respectively. In section \ref{sec:mixtures}, we describe how the set of cross sections, presented in section \ref{sec:results}, can be used to compute the effective ionization Townsend coefficient in gas mixtures containing C$_{3}$H$_{2}$F$_{4}$, potentially applicable for RPCs. Finally, conclusions are drawn in section \ref{sec:conclusion}.

\section{Numerical calculation techniques}\label{sec:techniques}

Solving the Boltzmann transport equation allows for the calculation of electron transport coefficients and reaction rate coefficients, starting from a given set of electron collision cross sections. This work is focused on the evaluation of the drift velocity $v_{drift}$, the longitudinal diffusion coefficient $D_{L}$, and the effective ionization rate coefficient $k_{eff}$ in order to compare them with the experimental data \cite{alise2017proceedings, alise2016paper}, available in the literature at present.

Two different numerical calculation techniques are used in this work to obtain $v_{drift}$, $D_{L}$ and $k_{eff}$ as a function of the reduced electric field $E/N$, where $E$ is the electric field and $N$ is the gas density. The first numerical technique solves the Boltzmann transport equation using the so-called two-term approximation (TTA) \cite{kumar1984physics, robson1986velocity, morgan1990elendif, hagelaar2005solving} while the second one is a Monte Carlo (MC) code \cite{fraser1986monte, biagi1999monte, rabie2016methes}. For our purposes we use the software BOLSIG+ \cite{hagelaar2005solving}, which exploits the TTA technique to calculate electron transport coefficients and reaction rate coefficients, whereas the MC code, used in this work, is METHES \cite{rabie2016methes}. Both programs have been validated by their developers with a large number of different gases and are widely used by the plasma modelling community \cite{hagelaar2005solving, rabie2016methes}. All input settings of BOLSIG+ and METHES are summarized in the appendix. These settings are used for all the results presented in this paper.

For a given set of electron collision cross sections, BOLSIG+ returns solutions of the Boltzmann transport equation much faster than METHES because of the different numerical techniques implemented. With the input settings mentioned in the appendix, the average computing time of BOLSIG+ is of the order of milliseconds while a METHES simulation lasts more than tens of minutes for each value of $E/N$ on a modern four-core PC (2.6 GHz CPU) \cite{hagelaar2005solving, rabie2016methes}. On the contrary, the TTA technique generally returns more approximate calculations than the MC technique, especially under specific conditions. Indeed, TTA calculations usually become less accurate at high values of $E/N$ where the electron energy distribution is not only determined by the elastic scattering but is also affected by the numerous excitation processes, which are energetically accessible \cite{sasic2013scattering, dupljanin2010transport}. Thus, MC techniques turn out to be more precise as their accuracy only depends on the statistical sampling during the MC simulation \cite{rabie2016methes, dupljanin2010transport}. In this work, all METHES calculations are performed with $10^{4}$ electrons for a minimum number of real collisions of $5\times10^{7}$. The accuracy of TTA and MC techniques, such as those implemented in BOLSIG+ and METHES respectively, is not further discussed in this paper, however more details are widely analyzed in several studies \cite{dupljanin2010transport, crompton1994benchmark, robson1997electron, petrovic2009measurement}.

\section{Methodology}\label{sec:methodology}

An iterative procedure has been implemented to obtain the scattering cross sections of electrons in C$_{3}$H$_{2}$F$_{4}$ from the existing measurements of electron transport coefficients and reaction rate coefficients. The method is a well-known technique described in several publications \cite{sasic2013scattering, crompton1994benchmark, robson1997electron}. The procedure consists of adjusting the cross sections of an initial set with the aim to improve progressively the agreement between the calculated swarm parameters and the experimental data. Fitting of the swarm parameters, obtained by the initial set of cross sections, within the experimental uncertainties, is usually achieved after a large number of iterations. For this reason, BOLSIG+ is here used to ease the iterative modifications of the cross sections because its calculations need a very short computing time for each iteration, whereas METHES is used at the end of the iterations to obtain accurate values of $v_{drift}$, $D_{L}$ and $k_{eff}$ as a function of $E/N$. In this paper, our conclusions are based on the METHES results.

One of the limitations of this iterative procedure is the non-uniqueness of electron collision cross sections. This can be solved by using direct measurements of electron collision cross sections or it may be mitigated thanks to theoretical calculations of some cross sections. Beam experiments are especially useful to provide additional information, such as the rates of some specific processes. However, this is not the case of the currently available experimental data in C$_{3}$H$_{2}$F$_{4}$ because they are obtained in a pulsed Townsend experiment by the analysis of signals in the electrodes \cite{alise2016paper}. Another way to reach the uniqueness of the final cross sections concerns the implementation of an iterative procedure by fitting experimental data in pure C$_{3}$H$_{2}$F$_{4}$ and also in mixtures of this gas. This methodology is generally carried out when experimental data are only available from swarm experiments as in the case of C$_{3}$H$_{2}$F$_{4}$. For this reason, we have obtained the electron collision cross sections in C$_{3}$H$_{2}$F$_{4}$ by fitting the electron transport coefficients and reaction rate coefficients in pure C$_{3}$H$_{2}$F$_{4}$ and also in gas mixtures of C$_{3}$H$_{2}$F$_{4}$ and CO$_{2}$ in different concentrations. It is important to highlight that gas mixtures including Ar are commonly taken into account to avoid the non-uniqueness of the cross sections \cite{crompton1994benchmark, sasic2013scattering}. However, we selected gas mixtures of C$_{3}$H$_{2}$F$_{4}$ and CO$_{2}$ for two reasons. Firstly, there are hints that additional Penning ionizations occur in Ar-based gas mixtures with the addition of C$_{3}$H$_{2}$F$_{4}$, as suggested by Chachereau et al. \cite{alise2016paper}, so the ionization rate can be altered by this process. Secondly, several ongoing studies aim to replace the standard mixtures in RPCs, which have high GWP values, with more environmental-friendly gas mixtures containing C$_{3}$H$_{2}$F$_{4}$ and CO$_{2}$ \cite{abbrescia2016preliminary, bianchi2019characterization, rigoletti2020studies, proto2022new, rigoletti2023studies}. Therefore, there is a growing interest in a reliable set of electron collision cross sections validated for gas mixtures of C$_{3}$H$_{2}$F$_{4}$ and CO$_{2}$ to reproduce accurately the electron transport parameters and reaction rates that play a crucial role in the signal formation in RPCs.

The set of electron collision cross sections in C$_{2}$H$_{2}$F$_{4}$, derived by {\v{S}}a{\v{s}}i{\'c} et al. \cite{sasic2010measurements}, is used as initial set in our iterative procedure. Indeed, the chemical structures of C$_{2}$H$_{2}$F$_{4}$ and C$_{3}$H$_{2}$F$_{4}$ are very similar, except for only one carbon atom, therefore it may be reasonable to assume that electron collision cross sections of both gases might be quite similar. As next step, the initial ionization cross section was shifted from $\sim$10 eV to 7.5 eV because previous simulations suggest that the ionization energy of C$_{3}$H$_{2}$F$_{4}$ is $\sim$3 eV lower than that of C$_{2}$H$_{2}$F$_{4}$ \cite{benussi2015properties}. Since the methodology involves adjusting the electron collision cross sections to match the measured electron swarm parameters, the elastic cross section of the initial set has been repeatedly adjusted by random variations of its magnitude in different energy ranges until the calculated values of drift velocity in pure C$_{3}$H$_{2}$F$_{4}$ fell within the experimental uncertainties. Moreover, adjustments were made to the magnitude of excitation cross sections to improve the agreement with the measured values of $v_{drift}$ as a function of $E/N$ in gas mixtures of C$_{3}$H$_{2}$F$_{4}$ and CO$_{2}$, in addition to those measured in pure C$_{3}$H$_{2}$F$_{4}$.

The attachment cross section of the initial set, corresponding to the electron attachment in C$_{2}$H$_{2}$F$_{4}$, cannot be used in the iterative procedure because there is experimental evidence indicating that electron capture in C$_{3}$H$_{2}$F$_{4}$ occurs through two distinct processes. Indeed, the effective ionization rate coefficient $k_{eff}$ in pure C$_{3}$H$_{2}$F$_{4}$ decreases when the gas pressure increases, as measured by Chachereau et al. \cite{alise2016paper}. This pressure dependence indicates the presence of three-body electron attachment processes \cite{alise2016paper, christophorou1984electrons}. In particular, the stable capture of electrons in C$_{3}$H$_{2}$F$_{4}$ can occur by dissociative attachment processes or collisional stabilization processes, described as follows:
\[e^{-} + C_{3}H_{2}F_{4} \rightarrow A^{-} + B \,\,\,\, \textrm{(dissociative attachment)}\]
\[(C_{3}H_{2}F_{4})^{*} + C_{3}H_{2}F_{4} \rightarrow (C_{3}H_{2}F_{4})^{-} + C_{3}H_{2}F_{4} \,\,\,\, \textrm{(collisional stabilization)}\]
where C$_{3}$H$_{2}$F$_{4}$, (C$_{3}$H$_{2}$F$_{4}$)$^{*}$ and (C$_{3}$H$_{2}$F$_{4}$)$^{-}$ indicate the neutral, excited and negative state of a molecule of C$_{3}$H$_{2}$F$_{4}$, respectively, whereas $e^{-}$ is the incident electron and $A^{-}$ and $B$ are the negative and neutral fragments of C$_{3}$H$_{2}$F$_{4}$ after its dissociation. In the first process, an incident electron $e^{-}$ interacts with a single neutral molecule of C$_{3}$H$_{2}$F$_{4}$, causing its dissociation into the negative fragment $A^{-}$ and the neutral fragment $B$. In the second process, an electron can be attached through the collision between the excited state of C$_{3}$H$_{2}$F$_{4}$, owing to a previous interaction with the incident electron, and a neutral molecule of C$_{3}$H$_{2}$F$_{4}$. This last type of electron capture, which involves three different bodies (one electron and two molecules of C$_{3}$H$_{2}$F$_{4}$), is called collision stabilization process and its probability depends on the collision rate of C$_{3}$H$_{2}$F$_{4}$ molecules, thereby influenced by the gas density $N$. For this reason, measurements of $k_{eff}$ are affected by the pressure at which data are taken. Similarly to the Bloch-Bradbury mechanism in O$_{2}$ \cite{bloch1935mechanism, aleksandrov1988three}, a simple kinetic model in the case of C$_{3}$H$_{2}$F$_{4}$ has been proposed by Chachereau et al. with the aim to reproduce the pressure dependence of effective ionization rate coefficient in this gas \cite{alise2016paper}.

As previously mentioned, the attachment cross section of the initial set, related to C$_{2}$H$_{2}$F$_{4}$, cannot describe both processes of electron capture that occur in C$_{3}$H$_{2}$F$_{4}$, so it was substituted by two different attachment cross sections, whose energy-dependence has been approximated as two parabolas with negative concavity as a first step. Since electrons that dissociate a gas molecule need usually more energy than electrons that are captured by three-body attachment, the initial cross section of the dissociation attachment was centered close to the ionization threshold, whereas the mean value of the other cross section, related to the three-body electron attachment, was chosen at 3 eV. Indeed, after several attempts at different energy values, the value of 3 eV turned out to be a good compromise to reproduce in first approximation the trend of effective ionization rate coefficient as a function of $E/N$ and gas pressure. Thanks to the same iterative procedure, described above, the final attachment cross sections were obtained by comparing the calculations and measurements of effective ionization rate coefficient as a function of $E/N$ in pure C$_{3}$H$_{2}$F$_{4}$ for different values of pressure.

\begin{figure}[ht]
    \centering
    \includegraphics[width=0.55\textwidth]{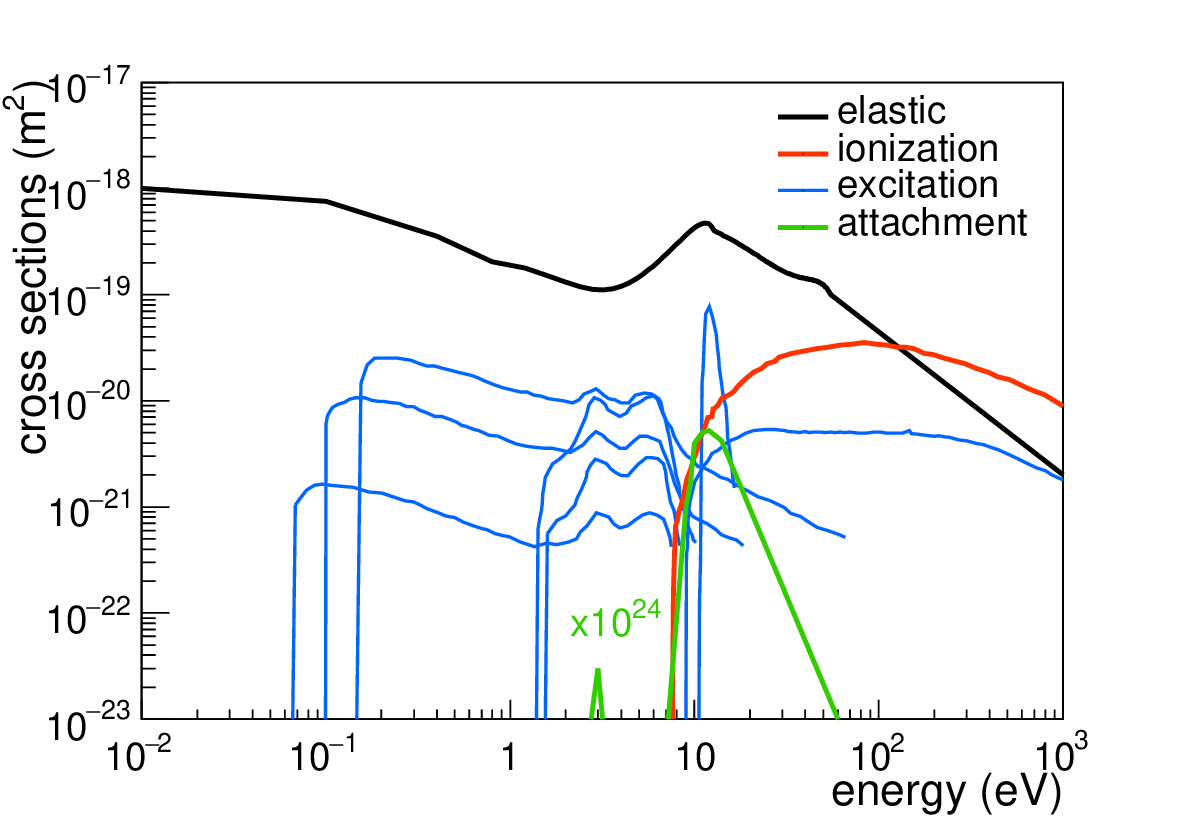}
    \caption{Final set of electron collision cross sections in C$_{3}$H$_{2}$F$_{4}$.}
    \label{fig:plot1}
\end{figure}

The final set of electron collision cross sections for C$_{3}$H$_{2}$F$_{4}$, obtained by the iterative procedure, is shown in figure \ref{fig:plot1}. The three-body attachment cross section centered at $\sim$3 eV is normalized by the gas density $N$.

\section{Results and discussion}\label{sec:results}
In this section, we present a systematic comparison between the electron swarm parameters calculated from the electron collision cross sections of C$_{3}$H$_{2}$F$_{4}$, shown in figure \ref{fig:plot1}, and the corresponding experimental data. The comparative study is especially focused on $v_{drift}$ and $k_{eff}$ as a function of $E/N$ at different values of gas pressure. The electron swarm parameters are calculated by BOLSIG+ and METHES, and then compared with the experimental data, measured by Chachereau et al., in pure C$_{3}$H$_{2}$F$_{4}$ and in gas mixtures of C$_{3}$H$_{2}$F$_{4}$/CO$_{2}$ and C$_{3}$H$_{2}$F$_{4}$/Ar. All experimental data are in the full paper \cite{alise2016paper} and in the conference paper \cite{alise2017proceedings}. Hereafter, we refer to BOLSIG+ with the notation TTA to highlight that numerical solutions of the Boltzmann transport equation are determined by the two-term approximation. On the contrary, solutions obtained through METHES are labeled as MC as they are calculated using the Monte Carlo method.

\subsection{Pure C$_{3}$H$_{2}$F$_{4}$}\label{sec:results1}
\begin{figure}[ht]
    \centering
    \includegraphics[width=0.55\textwidth]{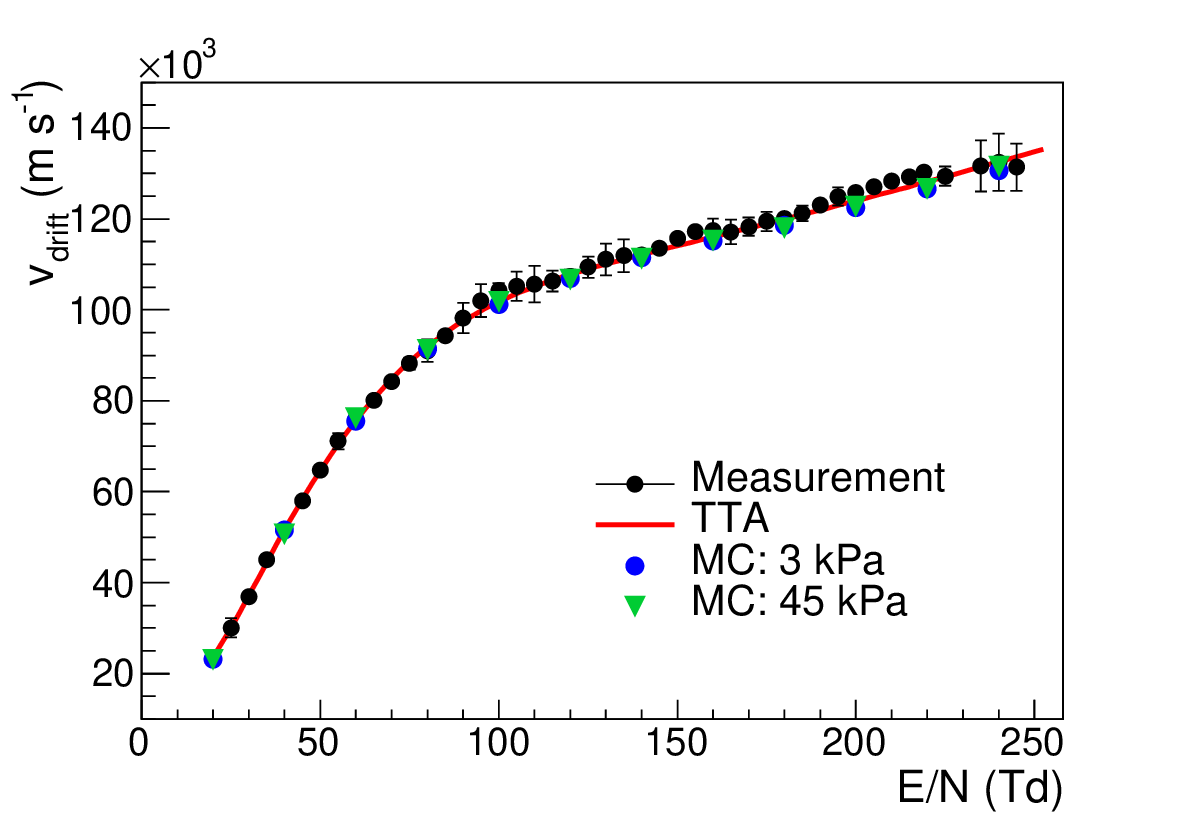}
    \caption{Electron drift velocity as a function of $E/N$ in pure C$_{3}$H$_{2}$F$_{4}$. TTA and MC results are compared with the measurements.}
    \label{fig:plot2}
\end{figure}

Values of $v_{drift}$ as a function of $E/N$ are measured at pressures ranging from 3 to 45 kPa and are found to be independent of gas pressure in pure C$_{3}$H$_{2}$F$_{4}$ \cite{alise2016paper}. Figure \ref{fig:plot2} shows measurements of $v_{drift}$ along with TTA and MC calculations, based on our electron collision cross sections of C$_{3}$H$_{2}$F$_{4}$ in figure \ref{fig:plot1}. Both TTA and MC calculations of $v_{drift}$ are in good agreement with experimental data. Indeed, the relative differences between calculations and measurements are lower than 3\% at both 3 kPa and 45 kPa across the entire range of $E/N$ investigated. These two pressure values represent the minimum and maximum pressures at which experimental data are available. Since no significant differences are observed in the TTA results of $v_{drift}$ at pressures ranging from 3 kPa to 45 kPa, the average of TTA calculations at 3 kPa and 45 kPa is plotted in figure \ref{fig:plot2} for each value of $E/N$.

\begin{figure}[ht]
    \centering
    \includegraphics[width=0.55\textwidth]{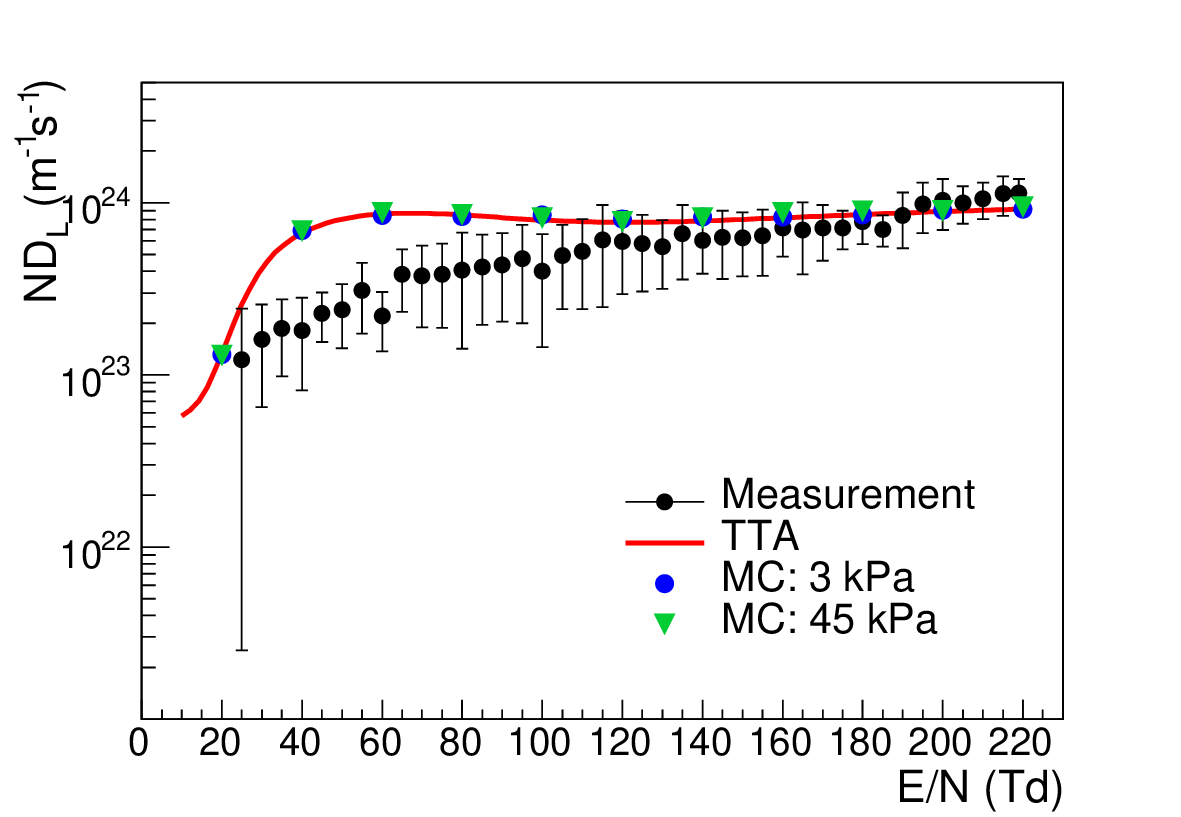}
    \caption{Density-normalized longitudinal diffusion coefficient as a function of $E/N$ in pure C$_{3}$H$_{2}$F$_{4}$. TTA and MC results are compared with the measurements.}
    \label{fig:plot3}
\end{figure}

Figure \ref{fig:plot3} shows the density-normalized longitudinal diffusion coefficient $ND_{L}$ as a function of $E/N$ in pure C$_{3}$H$_{2}$F$_{4}$. Similarly to $v_{drift}$, measurements of $ND_{L}$ are found to be independent of gas pressure in pure C$_{3}$H$_{2}$F$_{4}$ \cite{alise2016paper}. Since no significant differences are observed in the TTA results of $ND_{L}$ at pressures ranging from 3 kPa to 45 kPa, the average of TTA calculations at 3 kPa and 45 kPa is plotted in figure \ref{fig:plot3} for each value of $E/N$. Our set of cross sections accurately reproduces values of $ND_{L}$ only at high $E/N$. Indeed, TTA and MC calculations are within the experimental uncertainties for $E/N$ greater than 100 Td while differences between calculations and measurements become larger from 20 Td to 100 Td. This does not represent a fundamental issue if our set of electron collision cross sections is used to evaluate the performance of RPCs because they are usually operated in the range between $\sim$150 Td and $\sim$250 Td \cite{santonico1981development}. It is noted that 1 Td is equal to 10$^{-21}$ Vm$^{2}$.

\begin{figure}[ht]
    \centering
    \includegraphics[width=0.55\textwidth]{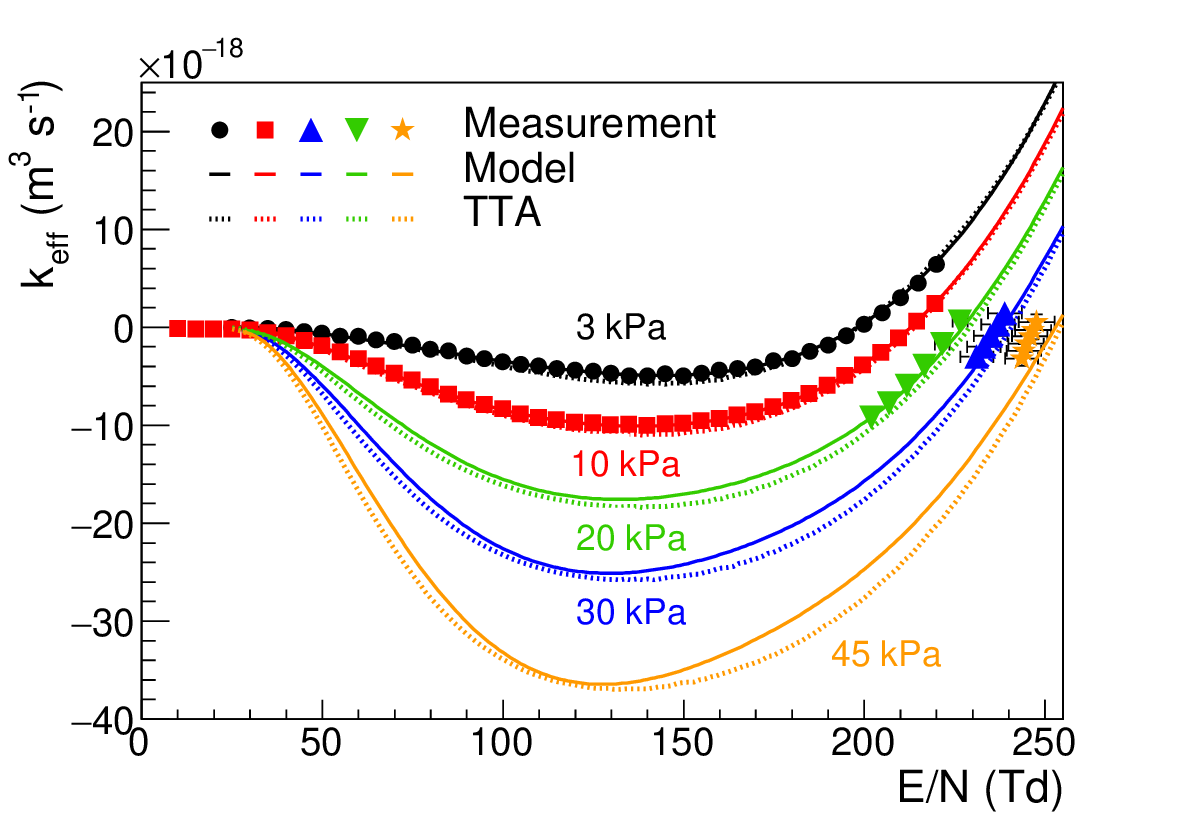}
    \caption{Effective ionization rate coefficient as a function of $E/N$ in pure C$_{3}$H$_{2}$F$_{4}$. TTA results are compared with the measurements at 3, 10, 20, 30 and 45 kPa.}
    \label{fig:plot4}
\end{figure}

\begin{figure}[ht]
    \centering
    \includegraphics[width=0.55\textwidth]{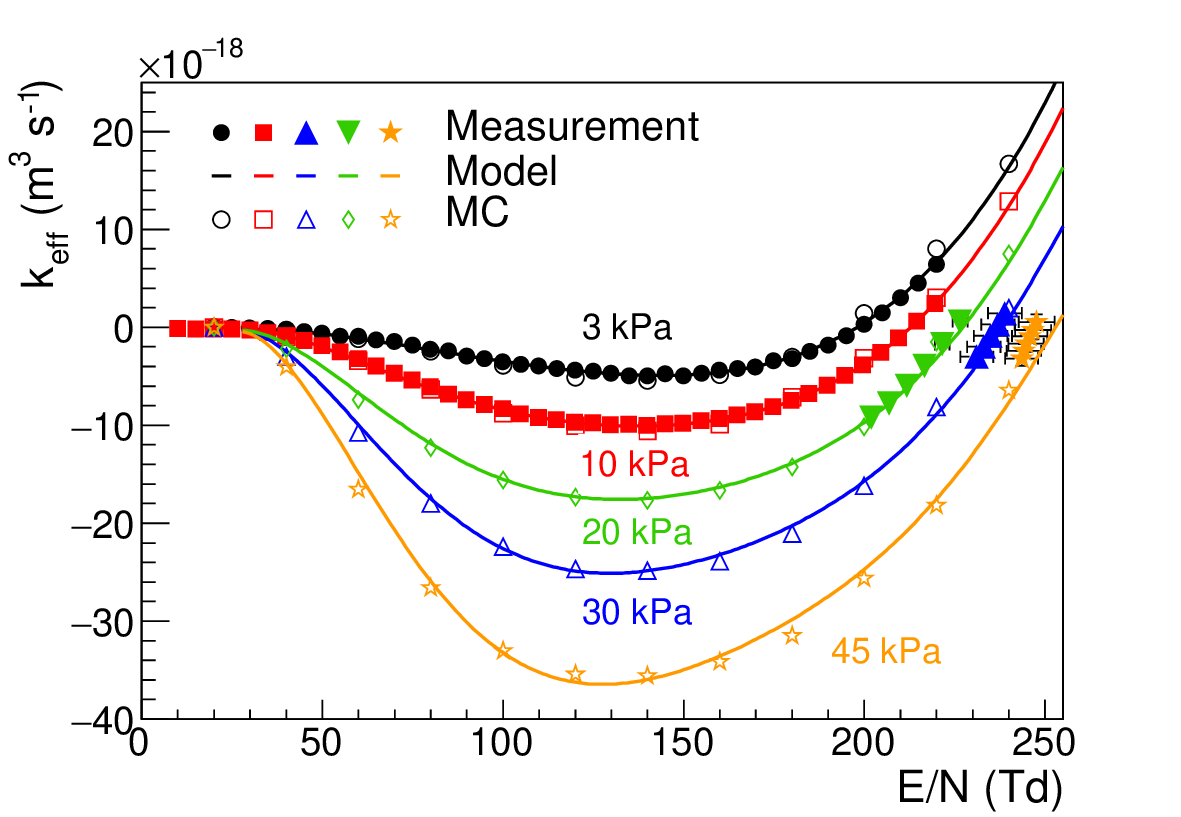}
    \caption{Effective ionization rate coefficient as a function of $E/N$ in pure C$_{3}$H$_{2}$F$_{4}$. MC results are compared with the measurements at 3, 10, 20, 30 and 45 kPa.}
    \label{fig:plot5}
\end{figure}

As previously mentioned, Chachereau et al. have developed a kinetic model to describe the pressure dependence of $k_{eff}$ in pure C$_{3}$H$_{2}$F$_{4}$. This model accurately reproduces the experimental measurements. Figure \ref{fig:plot4} shows the measurements of $k_{eff}$ and the corresponding values predicted by the model as a function of $E/N$ at different pressures. Moreover, TTA calculations based on our set of electron collision cross sections are also reported in figure \ref{fig:plot4}. Differences between the TTA calculations of $k_{eff}$ and predictions of the model turn out to be generally larger if the pressure increases. This is not observed in the case of MC calculations that are in agreement with the model predictions, as shown in figure \ref{fig:plot5}. Indeed, the trend of $k_{eff}$ as a function of $E/N$ and its pressure dependence are accurately reproduced by MC calculations, utilizing our set of electron collision cross sections, with the maximum difference between calculations and model predictions being lower than $1.5\times10^{-18}$ m${^3}$s$^{-1}$. It is well recognized that the TTA technique may not accurately compute the exact values of electron swarm parameters, especially when excitation processes become comparable to elastic scattering \cite{crompton1994benchmark, sasic2013scattering}. In general, codes based on a multi-term approximation tend to have higher accuracy compared to those utilizing a two-term approximation \cite{crompton1994benchmark, petrovic2009measurement, robson1997electron, white2003classical}. Hereafter, the MC integration will be the sole numerical technique considered for the following comparisons between calculations and measurements of electron swarm parameters in gas mixtures of C$_{3}$H$_{2}$F$_{4}$/CO$_{2}$ and C$_{3}$H$_{2}$F$_{4}$/Ar.

\begin{figure}[ht]
    \centering
    \includegraphics[width=0.55\textwidth]{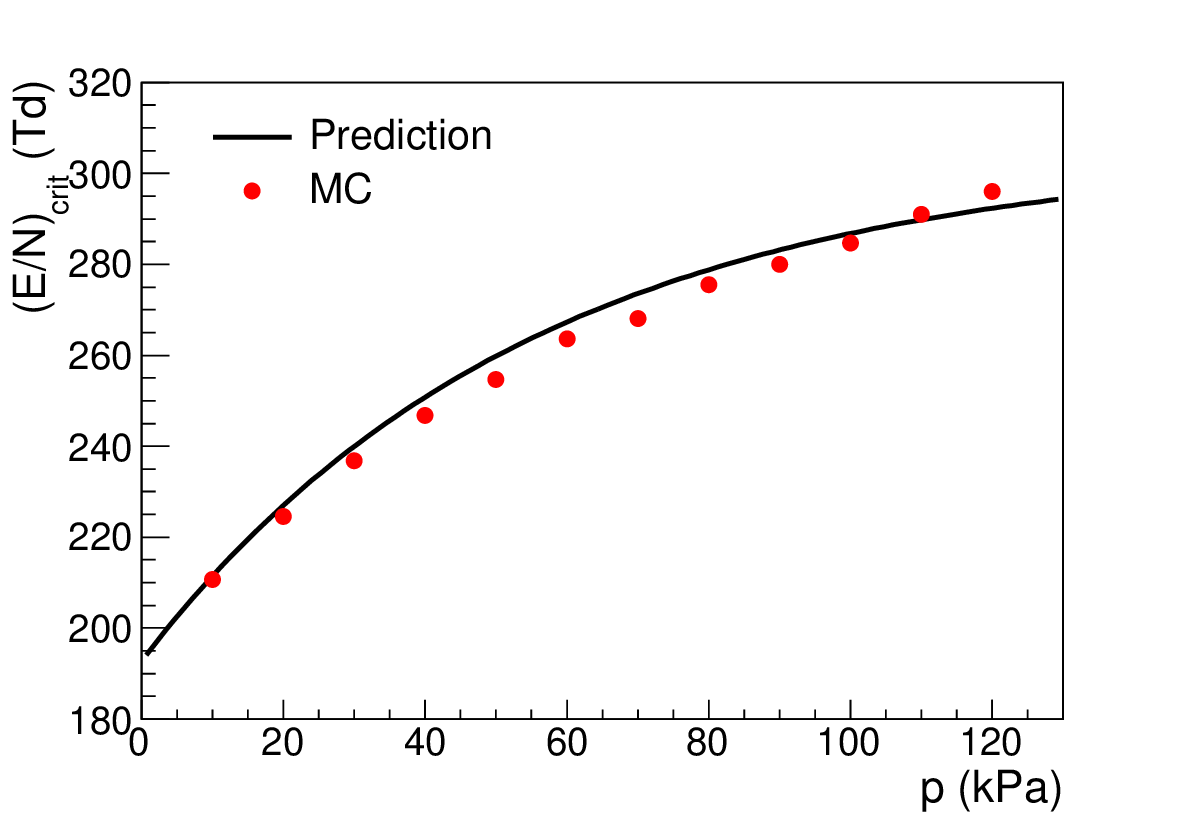}
    \caption{Reduced critical electric field strength as a function of the gas pressure in pure C$_{3}$H$_{2}$F$_{4}$. MC results, obtained using the electron collision cross sections in C$_{3}$H$_{2}$F$_{4}$, are compared with the prediction by Koch et al.}
    \label{fig:plot6}
\end{figure}

The reduced critical electric field strength $($E/N$)_{crit}$ is the value of $E/N$ at which $k_{eff}$ is zero. As a consequence, a discharge may occur and develop in the gas for values greater than  $($E/N$)_{crit}$ because ionization processes become predominant on electron attachments. For pure C$_{3}$H$_{2}$F$_{4}$, Koch et al. \cite{koch2015high} have estimated this quantity as follows:
\[(E/N)_{crit} = 305 \, \textrm{Td} \times \left(1 - \exp\left(-\frac{p + 55 \, \textrm{kPa}}{55 \, \textrm{kPa}}\right)\right) \]
where $p$ is the pressure of the gas. Figure \ref{fig:plot6} shows the comparison between the predictions by Koch et al. and the MC calculations based on our set of electron collision cross sections. As shown in figure \ref{fig:plot6}, a good agreement is achieved in the whole pressure range investigated. Relative errors between predictions by Koch et al. and our MC calculations are lower than 2\% for pressures ranging from 10 kPa to 120 kPa. This may confirm that scaling the three-body electron attachment cross section according to the gas density $N$ can be accurately applied for pressures greater than 45 kPa, for which no experimental data are currently available.

\subsection{Gas mixtures of C$_{3}$H$_{2}$F$_{4}$ and CO$_{2}$}\label{sec:results2}
\begin{figure}[ht]
    \centering
    \includegraphics[width=0.55\textwidth]{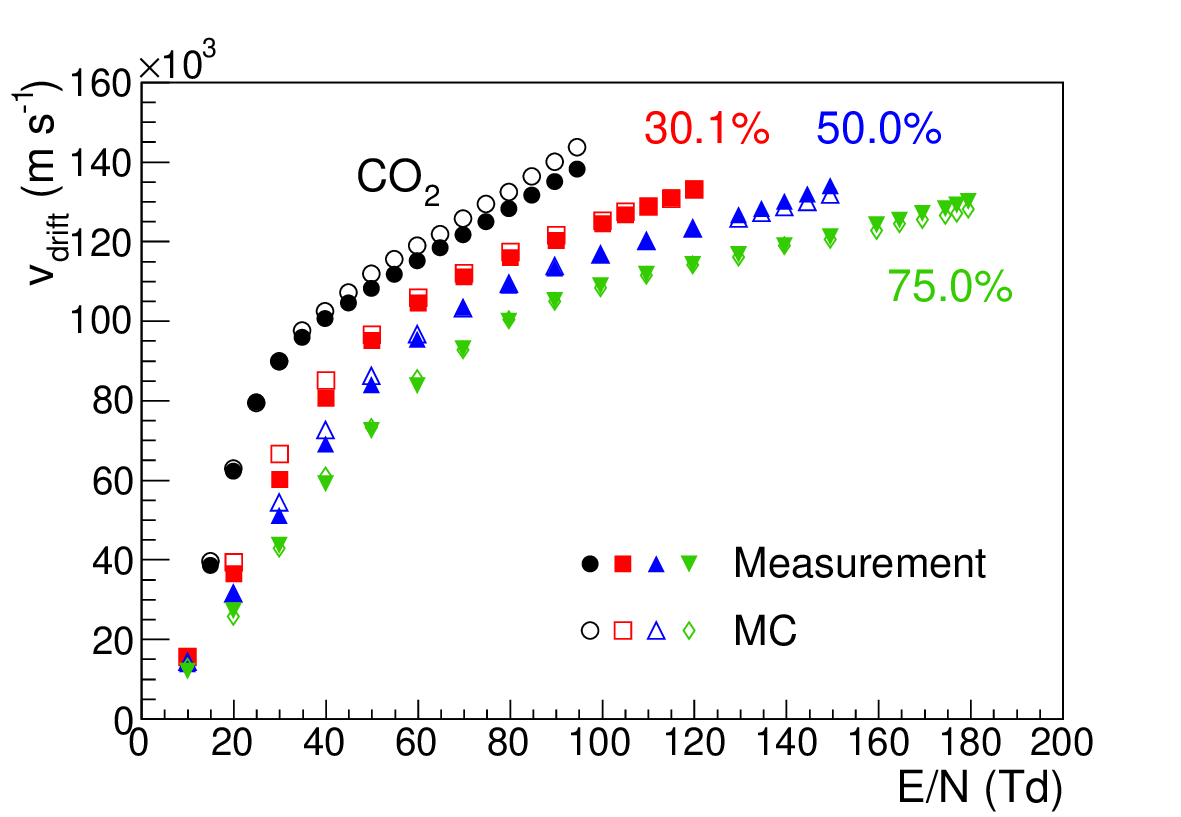}
    \caption{Drift velocity as a function of $E/N$ in pure CO$_{2}$ and in gas mixtures of 30.1\%, 50.0\% and 75\% C$_{3}$H$_{2}$F$_{4}$ with CO$_{2}$ at 10 kPa. MC results are compared with the measurements.}
    \label{fig:plot7}
\end{figure}

An important validation step to assess the accuracy of the derived electron collision cross sections involves comparing calculations with measurements of electron swarm parameters in binary mixtures. Figure \ref{fig:plot7} shows values of $v_{drift}$ as a function of $E/N$ for gas mixtures of C$_{3}$H$_{2}$F$_{4}$ and CO$_{2}$ in different concentrations at 10 kPa. The simulations are carried out for 30.1\%, 50.0\% and 75.0\% of C$_{3}$H$_{2}$F$_{4}$ in CO$_{2}$, whereas the cross sections of CO$_{2}$, used as input, are provided by Biagi \cite{biagi1999monte}, recently uploaded in the LXCat database \cite{projectlxcat} after an accurate review of its previous set. Calculations and measurements of $v_{drift}$ in mixtures of C$_{3}$H$_{2}$F$_{4}$ and CO$_{2}$ are generally in agreement. In particular, the relative differences are lower than 2\% in all gas mixtures of C$_{3}$H$_{2}$F$_{4}$/CO$_{2}$ considered, except for certain values of $E/N$ ranging from 20 Td to 50 Td, where the relative errors are approximately 10\%.

\begin{figure}[ht]
    \centering
    \includegraphics[width=0.55\textwidth]{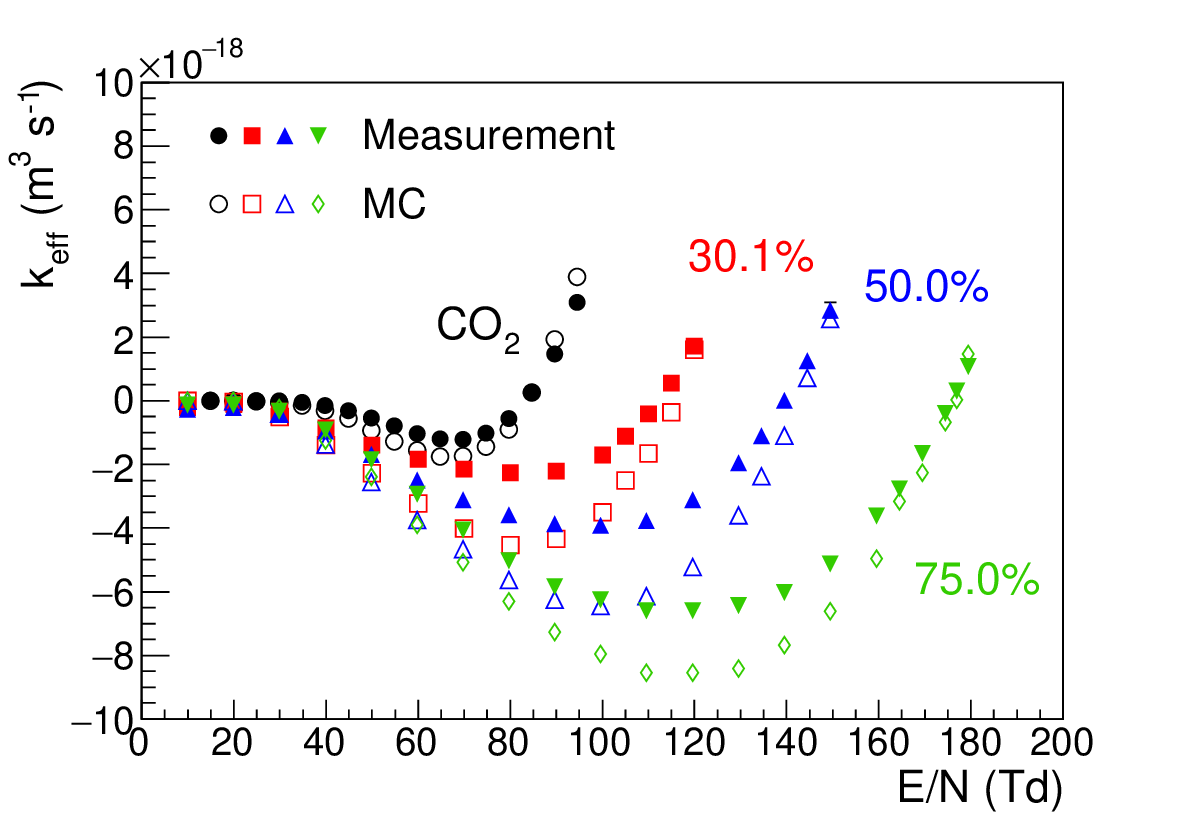}
    \caption{Effective ionization rate coefficient as a function of $E/N$ in pure CO$_{2}$ and in gas mixtures of 30.1\%, 50.0\% and 75\% C$_{3}$H$_{2}$F$_{4}$ with CO$_{2}$ at 10 kPa. MC results are compared with the measurements.}
    \label{fig:plot8}
\end{figure}

Values of $k_{eff}$ as a function of $E/N$ in gas mixtures of C$_{3}$H$_{2}$F$_{4}$ and CO$_{2}$ at 10 kPa are reported in figure \ref{fig:plot8}. Both results from calculations and measurements closely follow the same trends in all gas mixtures considered. However, calculations consistently yield lower values than measurements across the entire range of $E/N$ with a maximum difference of $\sim$$2.5 \times 10^{-18}$ m$^{3}$s$^{-1}$. Nevertheless, values of $($E/N$)_{crit}$ in these gas mixtures are determined with a relative error lower than 3\%.

\subsection{Ar-based gas mixtures with C$_{3}$H$_{2}$F$_{4}$}\label{sec:results3}
For completeness, we compare the calculations and measurements of $v_{drift}$ and $k_{eff}$ in Ar-based gas mixtures with addition of C$_{3}$H$_{2}$F$_{4}$. In particular, the study is focused on gas mixtures of 0.2\%, 0.9\%, and 2.8\% of C$_{3}$H$_{2}$F$_{4}$ in Ar at 10 kPa.

\begin{figure}[ht]
    \centering
    \includegraphics[width=0.55\textwidth]{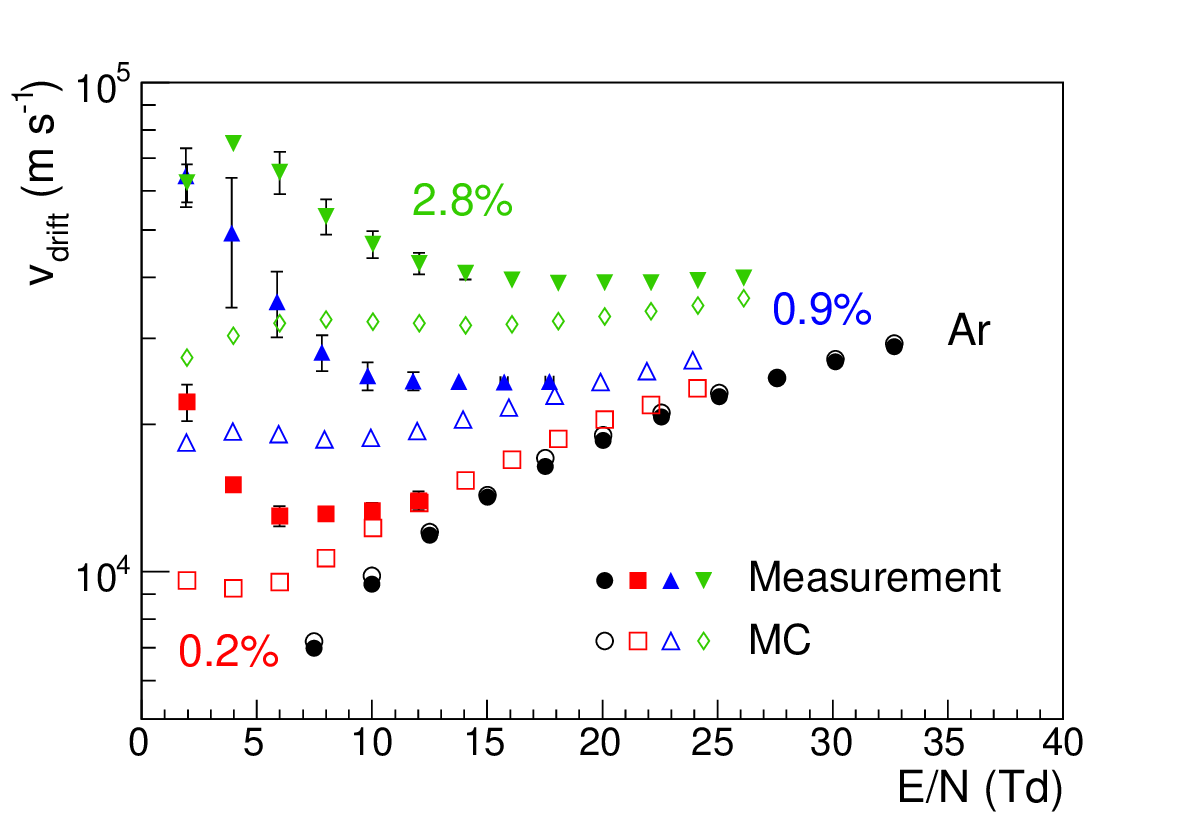}
    \caption{Drift velocity as a function of $E/N$ in pure Ar and Ar-based gas mixtures with 0.2\%, 0.9\%, and 2.8\% C$_{3}$H$_{2}$F$_{4}$ at 10 kPa. MC results are compared with the measurements.}
    \label{fig:plot9}
\end{figure}

Figure \ref{fig:plot9} shows values of $v_{drift}$ as a function of $E/N$ at 10 kPa. For the calculations, the cross sections of Ar, used as input, are provided by Biagi \cite{biagi1999monte}. A disagreement between MC results and measurements is evident, especially at low values of $E/N$, although the agreement improves at higher $E/N$. This suggests that further adjustments of the cross sections of C$_{3}$H$_{2}$F$_{4}$ in the low energy range may help reduce these differences. Nevertheless, the relative errors between calculations and measurements of $v_{drift}$ are lower than 30\% for $E/N$ values greater than 12 Td in all gas mixtures considered.

\begin{figure}[ht]
    \centering
    \includegraphics[width=0.55\textwidth]{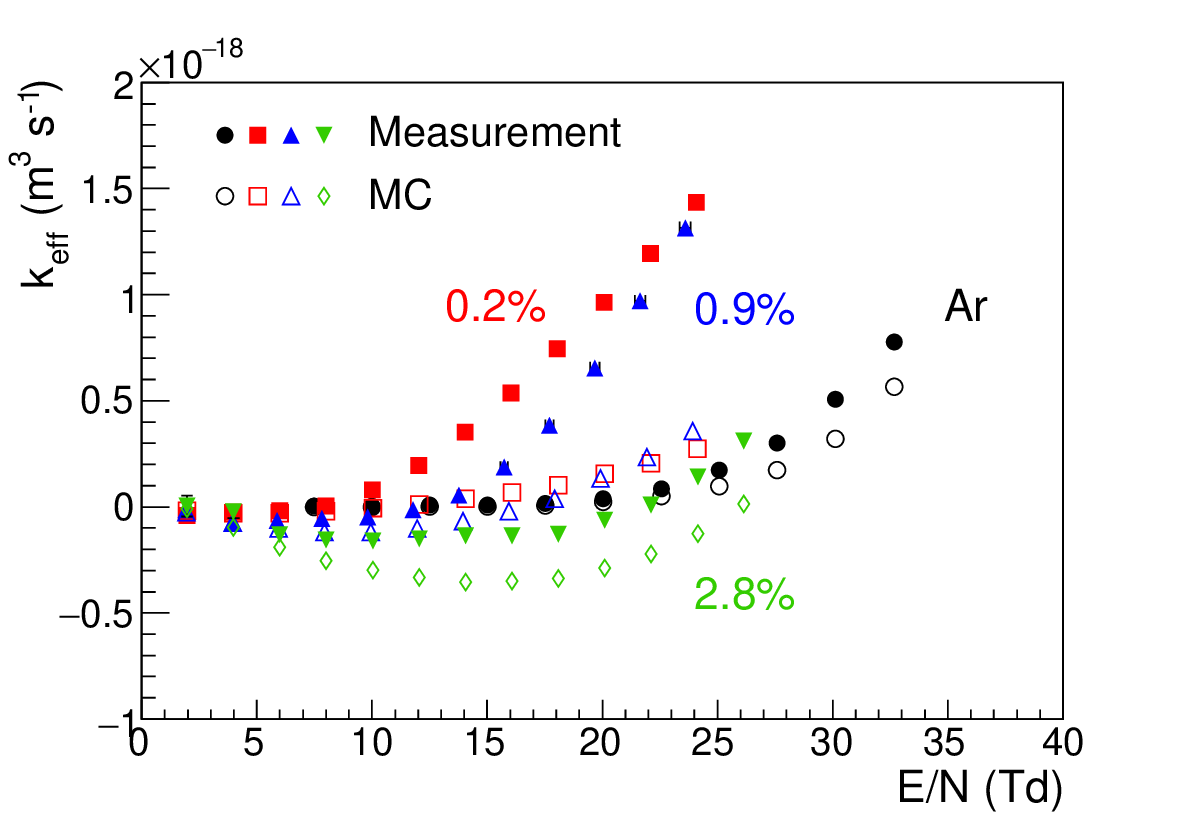}
    \caption{Effective ionization rate coefficient as a function of $E/N$ in pure Ar and Ar-based gas mixtures with 0.2\%, 0.9\%, and 2.8\% C$_{3}$H$_{2}$F$_{4}$ at 10 kPa. MC results are compared with the measurements.}
    \label{fig:plot10}
\end{figure}

Figure \ref{fig:plot10} shows the $k_{eff}$ values as a function of $E/N$ at 10 kPa. Significant differences between calculations and measurements are found in all Ar-based gas mixtures considered. This is expected due to the Penning effect between excited states of Ar and neutral molecules of C$_{3}$H$_{2}$F$_{4}$, as pointed out by Chachereau et al. \cite{alise2016paper}. Indeed, this additional ionization process is not accounted for in our set of electron collision cross sections, as it is specific to gas mixtures of C$_{3}$H$_{2}$F$_{4}$ in the presence of Ar.

\section{C$_{3}$H$_{2}$F$_{4}$-based gas mixtures for Resistive Plate Chambers}\label{sec:mixtures}

As demonstrated in the previous sections, the computation of drift velocity, effective ionization rate coefficient and longitudinal diffusion coefficient, based on our set of cross sections, are consistent with experimental measurements in both pure C$_{3}$H$_{2}$F$_{4}$ and gas mixtures containing C$_{3}$H$_{2}$F$_{4}$ and CO$_{2}$ in different percentages, especially for $E/N$ values higher than 100 Td. Some limitations are observed in the computation of those parameters in the presence of Ar. In light of these limitations, the set of cross sections proposed in this work finds applicability in various research fields, particularly in the substitution of standard gas mixtures in RPCs with more environmental-friendly mixtures based on C$_{3}$H$_{2}$F$_{4}$. Indeed, RPCs are typically operated at $E/N$ values higher than 100 Td and without the presence of Ar in the mixture.

As already evidenced in numerous studies \cite{abbrescia2016preliminary, bianchi2019characterization, rigoletti2020studies, proto2022new, rigoletti2023studies}, a feasible solution for RPCs to substitute their standard mixtures, containing mainly C$_{2}$H$_{2}$F$_{4}$, is the use of C$_{3}$H$_{2}$F$_{4}$. However, the direct substitution of C$_{2}$H$_{2}$F$_{4}$ with C$_{3}$H$_{2}$F$_{4}$ is not viable due to the resulting high operating voltages of RPCs. In this regard, it is expected that 2-mm-thick single-gap RPCs, operated solely with C$_{3}$H$_{2}$F$_{4}$ at room temperature and atmospheric pressure, would be efficient for voltages not lower than 14 kV \cite{bianchi2020siena}. Indeed, a recent study reports that the efficiency curve of RPCs is equal to zero until at least 14 kV, corresponding to approximately 260 Td. This is the maximum voltage tested during the study, to avoid irreversible damage to RPCs. A potential solution to address this issue is to replace C$_{2}$H$_{2}$F$_{4}$ with gas mixtures of C$_{3}$H$_{2}$F$_{4}$ and CO$_{2}$. Gas mixtures containing primarily CO$_{2}$ and a percentage ranging from 30\% to 50\% of C$_{3}$H$_{2}$F$_{4}$ are currently promising in ensuring the desired performance of RPCs without the need to operate at excessively high voltages.

\begin{figure}[ht]
    \centering
    \includegraphics[width=0.55\textwidth]{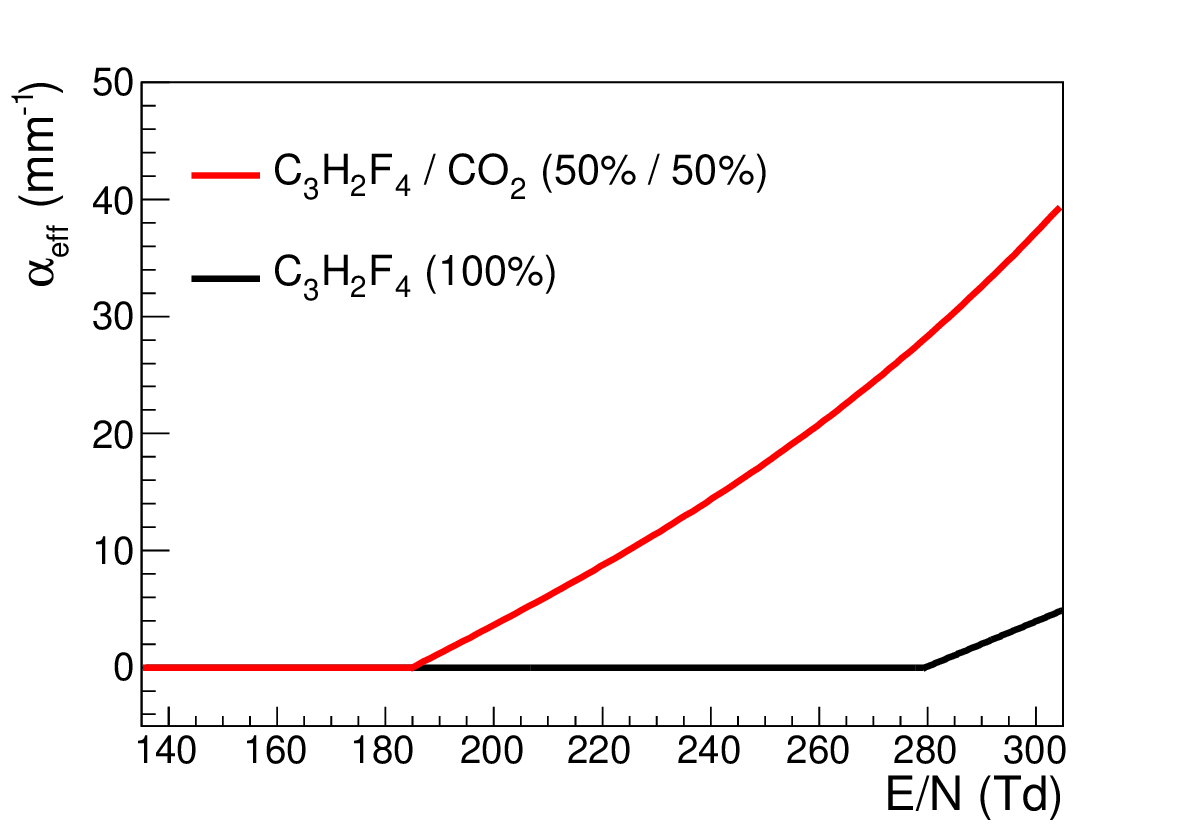}
    \caption{Effective ionization Townsend coefficient $\alpha_{eff}$ as a function of $E/N$ in pure C$_{3}$H$_{2}$F$_{4}$ and in a gas mixture containing 50\% C$_{3}$H$_{2}$F$_{4}$ and 50\% CO$_{2}$. Simulation results are obtained using the MATOQ code.}
    \label{fig:plot11}
\end{figure}

To highlight the effectiveness of our set of electron collision cross sections in simulating RPCs with gas mixtures containing C$_{3}$H$_{2}$F$_{4}$, we calculated the effective ionization Townsend coefficient using our set of cross sections and the MATOQ code, a Monte Carlo simulation specifically developed for gas mixtures for RPCs \cite{bianchi2023matoq}. If this coefficient is greater than zero, ionization events occur more frequently than attachment events. When the effective Townsend ionization coefficient, with increasing the electric field, begins to deviate from zero, electron avalanches start to develop in the gas gaps of RPCs, thus providing an indication of the electric field value at which the efficiency of RPCs begins to deviate from zero, especially when the discrimination threshold of the front-end electronics is quite low. Figure \ref{fig:plot11} shows the effective ionization Townsend coefficient $\alpha_{eff}$ as a function of $E/N$ in pure C$_{3}$H$_{2}$F$_{4}$ and in a gas mixture containing 50\% C$_{3}$H$_{2}$F$_{4}$ and 50\% CO$_{2}$. As shown in figure \ref{fig:plot11}, in the case of pure C$_{3}$H$_{2}$F$_{4}$, the effective ionization Townsend coefficient begins to deviate from zero at around 280 Td, corresponding to about 15 kV in a 2-mm-thick single-gap RPC. The value of 15 kV is consistent with experimental measurements, where this type of RPC is found to be inefficient until a voltage of at least 14 kV \cite{bianchi2020siena}. In the case of a gas mixture with the same percentage of C$_{3}$H$_{2}$F$_{4}$ and CO$_{2}$, the effective ionization Townsend coefficient begins to deviate from zero at around 187 Td. This value, corresponding to approximately 10.0 kV applied across the electrodes of a 2-mm-thick single-gap RPC, provides a good indication of the measured voltage at which the efficiency begins to deviate from zero. In fact, a recent study reports that RPCs of the same type operating with a mixture of 50\% C$_{3}$H$_{2}$F$_{4}$ and 50\% CO$_{2}$ turn out to be completely inefficient until 10.3 kV \cite{bianchiRPC2020}. The slight difference between the predicted value of 10.0 kV and the measured value of 10.3 kV may be attributed to the discrimination threshold of approximately 130 fC used in the front-end electronics for data acquisition \cite{bianchiRPC2020}.

Further demonstrating the effectiveness of our electron collision cross sections in simulating RPCs with gas mixtures that contain C$_{3}$H$_{2}$F$_{4}$ goes beyond the scope of this paper. The reader is referred to consult previous studies on RPCs with single gaps of 2 mm \cite{bianchiRPC2020} and 0.1 mm \cite{bianchi2023matoq}, operated with C$_{3}$H$_{2}$F$_{4}$-based gas mixtures.

\section{Conclusions}\label{sec:conclusion}
In this study, we implemented an iterative procedure to obtain a complete set of electron collision cross sections in tetrafluoropropene C$_{3}$H$_{2}$F$_{4}$ with the trade name HFO1234ze(E). The technique, widely used for numerous gas molecules including those with multiple fluorine atoms, consists of adjusting the cross sections of an initial set to progressively improve the agreement between the calculated swarm parameters and the experimental measurements. Using the electron collision cross sections proposed in this work, the computed drift velocity and effective ionization rate coefficient as a function of $E/N$ are in agreement with experimental measurements, both in pure C$_{3}$H$_{2}$F$_{4}$ and in gas mixtures containing varying percentages of C$_{3}$H$_{2}$F$_{4}$ and CO$_{2}$. Our set of electron collision cross sections also enables an accurate prediction of the reduced critical electric field strength as a function of pressure. However, discrepancies are observed in the agreement between simulated and measured values of the longitudinal diffusion coefficient in pure C$_{3}$H$_{2}$F$_{4}$ for $E/N$ values lower than 100 Td. This suggests that further adjustments to the current set may be needed to enhance the agreement with the experimental data for $E/N$ values lower than 100 Td. Direct measurements of at least some electron collision cross sections of C$_{3}$H$_{2}$F$_{4}$ would be advisable. Another limitation that has been found concerns the simulation of drift velocity and effective ionization rate coefficient in Ar-based gas mixtures with the addition of C$_{3}$H$_{2}$F$_{4}$. This is expected due to the Penning effect between Ar and C$_{3}$H$_{2}$F$_{4}$, as already reported in the literature. Nevertheless, the current version of electron collision cross sections presented here provides valuable insights for numerous studies, especially regarding the substitution of standard mixtures in RPCs with more environmental-friendly gas mixtures containing C$_{3}$H$_{2}$F$_{4}$. In this context, the current set of electron collision cross sections is applicable in simulating RPCs, as they are typically operated at $E/N$ values higher than 100 Td, and Ar is usually not included in their gas mixtures. To highlight the effectiveness of our set of electron collision cross sections in simulating electron swarm parameters in gas mixtures for RPCs, some simulations have been presented in this work. These simulation results concerning the effective ionization Townsend coefficient, obtained in both pure C$_{3}$H$_{2}$F$_{4}$ and a gas mixture with 50\% C$_{3}$H$_{2}$F$_{4}$ and 50\% CO$_{2}$, are consistent with experimental measurements.

\section*{Acknowledgments}
We thank Dr. Alise Chachereau and Prof. Dr. Christian Franck for sharing their data with us.

\section*{Appendix}
For the TTA technique used in this work, the settings of BOLSIG+ code (version 3/2016) are summarized in table \ref{settings_BOLSIG}. The type of electron growth is set as \textit{temporal}, except for the calculation of $ND_{L}$ where the setting has to be \textit{grad-n expansion}. All other settings, not specified in this appendix, has to be considered equal to the default settings of BOLSIG+. 
\begin{table}[ht]
\begin{center}
 \begin{tabular}{l l} 
 \hline
 Variable & Value \\ 
 \hline\hline
 Gas temperature & 300.0 K  \\ 
 \hline
 Maximum number of iterations & 200.0 eV \\
 \hline
 Maximum energy & 1000 \\
 \hline
\end{tabular}
\caption{Settings of BOLSIG+}
\label{settings_BOLSIG}
\end{center}
\end{table}

Concerning the MC integration technique, the settings of METHES (version 3, developed in June 2016) are reported in table \ref{settings_METHES}, where $N_{0}$ is the initial number of electrons that remains constant in the simulation, $col_{equ}$ is the minimum number of collisions before reaching the steady state, and $w_{err}$ and $DN_{err}$ are the error tolerance for the flux drift velocity and flux diffusion coefficient, respectively. All other settings, not specified in this appendix, are considered equal to the default settings of METHES. All transport coefficients obtained by METHES and presented in this study are the \textit{flux values} \cite{rabie2016methes}.
\begin{table}[ht]
\begin{center}
 \begin{tabular}{l l} 
 \hline
 Variable & Value \\ 
 \hline\hline
 Gas temperature&300.0 K\\
 \hline
 $N_{0}$&$10^{4}$\\
 \hline
 $w_{err}$&0.5\%\\
 \hline
 $DN_{err}$&1\%\\
 \hline
 $col_{equ}$&$5 \times 10^{7}$\\
 \hline
\end{tabular}
\caption{Settings of METHES}
\label{settings_METHES}
\end{center}
\end{table}

Sets of electron collision cross sections in Ar and CO$_{2}$, used in the calculations of this study, were obtained from the database by Biagi \cite{biagi1999monte}, accessible on the LXCat website \cite{projectlxcat}. Electron collision cross sections in Ar were uploaded on October 20th, 2011, while those in CO$_{2}$ were uploaded on October 2nd, 2019.

\section*{\label{sec:level14}DATA AVAILABILITY}
The data that support the findings of this study are available from the authors upon reasonable request.

%%\section*{\label{sec:level6}ACKNOWLEDGEMENTS}

%%\section*{\label{sec:level7}DATA AVAILABILITY}

%%\section*{\label{sec:level8}AUTHOR CONTRIBUTIONS}
% ATTENZIONE: SEZIONE DA COMPLETARE!

%%\section*{\label{sec:level9}COMPETING INTERESTS}
%%The authors declare no competing interests.

%%\section*{\label{sec:level10}ADDITIONAL INFORMATION}
%%Correspondence and requests for materials should be addressed to A.B.

% The \nocite command causes all entries in a bibliography to be printed out
% whether or not they are actually referenced in the text. This is appropriate
% for the sample file to show the different styles of references, but authors
% most likely will not want to use it.
%\nocite{*}

\renewcommand{\bibsection}{\subsection*{REFERENCES}}

\bibliography{apssamp}% Produces the bibliography via BibTeX.

%apsrev4-2.bst 2019-01-14 (MD) hand-edited version of apsrev4-1.bst
%Control: key (0)
%Control: author (8) initials jnrlst
%Control: editor formatted (1) identically to author
%Control: production of article title (0) allowed
%Control: page (0) single
%Control: year (1) truncated
%Control: production of eprint (0) enabled
\begin{thebibliography}{36}%
\makeatletter
\providecommand \@ifxundefined [1]{%
 \@ifx{#1\undefined}
}%
\providecommand \@ifnum [1]{%
 \ifnum #1\expandafter \@firstoftwo
 \else \expandafter \@secondoftwo
 \fi
}%
\providecommand \@ifx [1]{%
 \ifx #1\expandafter \@firstoftwo
 \else \expandafter \@secondoftwo
 \fi
}%
\providecommand \natexlab [1]{#1}%
\providecommand \enquote  [1]{``#1''}%
\providecommand \bibnamefont  [1]{#1}%
\providecommand \bibfnamefont [1]{#1}%
\providecommand \citenamefont [1]{#1}%
\providecommand \href@noop [0]{\@secondoftwo}%
\providecommand \href [0]{\begingroup \@sanitize@url \@href}%
\providecommand \@href[1]{\@@startlink{#1}\@@href}%
\providecommand \@@href[1]{\endgroup#1\@@endlink}%
\providecommand \@sanitize@url [0]{\catcode `\\12\catcode `\$12\catcode `\&12\catcode `\#12\catcode `\^12\catcode `\_12\catcode `\%12\relax}%
\providecommand \@@startlink[1]{}%
\providecommand \@@endlink[0]{}%
\providecommand \url  [0]{\begingroup\@sanitize@url \@url }%
\providecommand \@url [1]{\endgroup\@href {#1}{\urlprefix }}%
\providecommand \urlprefix  [0]{URL }%
\providecommand \Eprint [0]{\href }%
\providecommand \doibase [0]{https://doi.org/}%
\providecommand \selectlanguage [0]{\@gobble}%
\providecommand \bibinfo  [0]{\@secondoftwo}%
\providecommand \bibfield  [0]{\@secondoftwo}%
\providecommand \translation [1]{[#1]}%
\providecommand \BibitemOpen [0]{}%
\providecommand \bibitemStop [0]{}%
\providecommand \bibitemNoStop [0]{.\EOS\space}%
\providecommand \EOS [0]{\spacefactor3000\relax}%
\providecommand \BibitemShut  [1]{\csname bibitem#1\endcsname}%
\let\auto@bib@innerbib\@empty
%</preamble>
\bibitem [{\citenamefont {{Intergovernmental Panel on Climate Change (IPCC)}}(2014)}]{IPCC}%
  \BibitemOpen
  \bibfield  {author} {\bibinfo {author} {\bibnamefont {{Intergovernmental Panel on Climate Change (IPCC)}}},\ }\href@noop {} {\bibinfo {title} {Fifth assessment report}} (\bibinfo {year} {2014})\BibitemShut {NoStop}%
\bibitem [{\citenamefont {{The European Parliament and The Council}}()}]{EUregulation}%
  \BibitemOpen
  \bibfield  {author} {\bibinfo {author} {\bibnamefont {{The European Parliament and The Council}}},\ }\href@noop {} {\bibinfo {title} {{Regulation (EU) No 517/2014 on fluorinated greenhouse gases}}},\ \bibinfo {note} {{OJL 150 (2014-05-04), pp. 195--230}}\BibitemShut {NoStop}%
\bibitem [{\citenamefont {Santonico}\ and\ \citenamefont {Cardarelli}(1981)}]{santonico1981development}%
  \BibitemOpen
  \bibfield  {author} {\bibinfo {author} {\bibfnamefont {R.}~\bibnamefont {Santonico}}\ and\ \bibinfo {author} {\bibfnamefont {R.}~\bibnamefont {Cardarelli}},\ }\bibfield  {title} {\bibinfo {title} {Development of {R}esistive {P}late {C}ounters},\ }\href@noop {} {\bibfield  {journal} {\bibinfo  {journal} {Nuclear Instruments and Methods in Physics Research}\ }\textbf {\bibinfo {volume} {187}},\ \bibinfo {pages} {377} (\bibinfo {year} {1981})}\BibitemShut {NoStop}%
\bibitem [{\citenamefont {Abbrescia}\ \emph {et~al.}(2016)\citenamefont {Abbrescia}, \citenamefont {Van~Auwegem}, \citenamefont {Benussi}, \citenamefont {Bianco}, \citenamefont {Cauwenbergh}, \citenamefont {Ferrini}, \citenamefont {Muhammad}, \citenamefont {Passamonti}, \citenamefont {Pierluigi}, \citenamefont {Piccolo} \emph {et~al.}}]{abbrescia2016preliminary}%
  \BibitemOpen
  \bibfield  {author} {\bibinfo {author} {\bibfnamefont {M.}~\bibnamefont {Abbrescia}}, \bibinfo {author} {\bibfnamefont {P.}~\bibnamefont {Van~Auwegem}}, \bibinfo {author} {\bibfnamefont {L.}~\bibnamefont {Benussi}}, \bibinfo {author} {\bibfnamefont {S.}~\bibnamefont {Bianco}}, \bibinfo {author} {\bibfnamefont {S.}~\bibnamefont {Cauwenbergh}}, \bibinfo {author} {\bibfnamefont {M.}~\bibnamefont {Ferrini}}, \bibinfo {author} {\bibfnamefont {S.}~\bibnamefont {Muhammad}}, \bibinfo {author} {\bibfnamefont {L.}~\bibnamefont {Passamonti}}, \bibinfo {author} {\bibfnamefont {D.}~\bibnamefont {Pierluigi}}, \bibinfo {author} {\bibfnamefont {D.}~\bibnamefont {Piccolo}}, \emph {et~al.},\ }\bibfield  {title} {\bibinfo {title} {Preliminary results of {R}esistive {P}late {C}hambers operated with eco-friendly gas mixtures for application in the {CMS} experiment},\ }\href@noop {} {\bibfield  {journal} {\bibinfo  {journal} {Journal of Instrumentation}\ }\textbf {\bibinfo {volume} {11}}\bibinfo  {number} { (09)},\ \bibinfo
  {pages} {C09018}}\BibitemShut {NoStop}%
\bibitem [{\citenamefont {Bianchi}\ \emph {et~al.}(2019)\citenamefont {Bianchi}, \citenamefont {Delsanto}, \citenamefont {Dupieux}, \citenamefont {Ferretti}, \citenamefont {Gagliardi}, \citenamefont {Joly}, \citenamefont {Manen}, \citenamefont {Marchisone}, \citenamefont {Micheletti}, \citenamefont {Rosano} \emph {et~al.}}]{bianchi2019characterization}%
  \BibitemOpen
\bibfield  {number} {  }\bibfield  {author} {\bibinfo {author} {\bibfnamefont {A.}~\bibnamefont {Bianchi}}, \bibinfo {author} {\bibfnamefont {S.}~\bibnamefont {Delsanto}}, \bibinfo {author} {\bibfnamefont {P.}~\bibnamefont {Dupieux}}, \bibinfo {author} {\bibfnamefont {A.}~\bibnamefont {Ferretti}}, \bibinfo {author} {\bibfnamefont {M.}~\bibnamefont {Gagliardi}}, \bibinfo {author} {\bibfnamefont {B.}~\bibnamefont {Joly}}, \bibinfo {author} {\bibfnamefont {S.}~\bibnamefont {Manen}}, \bibinfo {author} {\bibfnamefont {M.}~\bibnamefont {Marchisone}}, \bibinfo {author} {\bibfnamefont {L.}~\bibnamefont {Micheletti}}, \bibinfo {author} {\bibfnamefont {A.}~\bibnamefont {Rosano}}, \emph {et~al.},\ }\bibfield  {title} {\bibinfo {title} {Characterization of tetrafluoropropene-based gas mixtures for the {R}esistive {P}late {C}hambers of the {ALICE} muon spectrometer},\ }\href@noop {} {\bibfield  {journal} {\bibinfo  {journal} {Journal of Instrumentation}\ }\textbf {\bibinfo {volume} {14}}\bibinfo  {number} { (11)},\
  \bibinfo {pages} {P11014}}\BibitemShut {NoStop}%
\bibitem [{\citenamefont {Rigoletti}\ \emph {et~al.}(2020)\citenamefont {Rigoletti}, \citenamefont {Aielli}, \citenamefont {Alberghi}, \citenamefont {Benussi}, \citenamefont {Bianchi}, \citenamefont {Bianco}, \citenamefont {Di~Stante}, \citenamefont {Boscherini}, \citenamefont {Bruni}, \citenamefont {Camarri} \emph {et~al.}}]{rigoletti2020studies}%
  \BibitemOpen
\bibfield  {number} {  }\bibfield  {author} {\bibinfo {author} {\bibfnamefont {G.}~\bibnamefont {Rigoletti}}, \bibinfo {author} {\bibfnamefont {G.}~\bibnamefont {Aielli}}, \bibinfo {author} {\bibfnamefont {G.}~\bibnamefont {Alberghi}}, \bibinfo {author} {\bibfnamefont {L.}~\bibnamefont {Benussi}}, \bibinfo {author} {\bibfnamefont {A.}~\bibnamefont {Bianchi}}, \bibinfo {author} {\bibfnamefont {S.}~\bibnamefont {Bianco}}, \bibinfo {author} {\bibfnamefont {L.}~\bibnamefont {Di~Stante}}, \bibinfo {author} {\bibfnamefont {D.}~\bibnamefont {Boscherini}}, \bibinfo {author} {\bibfnamefont {A.}~\bibnamefont {Bruni}}, \bibinfo {author} {\bibfnamefont {P.}~\bibnamefont {Camarri}}, \emph {et~al.},\ }\bibfield  {title} {\bibinfo {title} {Studies of {RPC} detector operation with eco-friendly gas mixtures under irradiation at the {CERN} {G}amma {I}rradiation {F}acility},\ }\href@noop {} {\bibfield  {journal} {\bibinfo  {journal} {POS Proceedings of Science}\ }\textbf {\bibinfo {volume} {364}},\ \bibinfo {pages} {164}
  (\bibinfo {year} {2020})}\BibitemShut {NoStop}%
\bibitem [{\citenamefont {Proto}\ \emph {et~al.}(2022)\citenamefont {Proto}, \citenamefont {Liberti}, \citenamefont {Santonico}, \citenamefont {Aielli}, \citenamefont {Camarri}, \citenamefont {Cardarelli}, \citenamefont {Di~Ciaccio}, \citenamefont {Di~Stante}, \citenamefont {Paoloni}, \citenamefont {Pastori} \emph {et~al.}}]{proto2022new}%
  \BibitemOpen
  \bibfield  {author} {\bibinfo {author} {\bibfnamefont {G.}~\bibnamefont {Proto}}, \bibinfo {author} {\bibfnamefont {B.}~\bibnamefont {Liberti}}, \bibinfo {author} {\bibfnamefont {R.}~\bibnamefont {Santonico}}, \bibinfo {author} {\bibfnamefont {G.}~\bibnamefont {Aielli}}, \bibinfo {author} {\bibfnamefont {P.}~\bibnamefont {Camarri}}, \bibinfo {author} {\bibfnamefont {R.}~\bibnamefont {Cardarelli}}, \bibinfo {author} {\bibfnamefont {A.}~\bibnamefont {Di~Ciaccio}}, \bibinfo {author} {\bibfnamefont {L.}~\bibnamefont {Di~Stante}}, \bibinfo {author} {\bibfnamefont {A.}~\bibnamefont {Paoloni}}, \bibinfo {author} {\bibfnamefont {E.}~\bibnamefont {Pastori}}, \emph {et~al.},\ }\bibfield  {title} {\bibinfo {title} {On a new environment-friendly gas mixture for {R}esistive {P}late {C}hambers},\ }\href@noop {} {\bibfield  {journal} {\bibinfo  {journal} {Journal of Instrumentation}\ }\textbf {\bibinfo {volume} {17}}\bibinfo  {number} { (05)},\ \bibinfo {pages} {P05005}}\BibitemShut {NoStop}%
\bibitem [{\citenamefont {Rigoletti}\ \emph {et~al.}(2023)\citenamefont {Rigoletti}, \citenamefont {Guida},\ and\ \citenamefont {Mandelli}}]{rigoletti2023studies}%
  \BibitemOpen
\bibfield  {number} {  }\bibfield  {author} {\bibinfo {author} {\bibfnamefont {G.}~\bibnamefont {Rigoletti}}, \bibinfo {author} {\bibfnamefont {R.}~\bibnamefont {Guida}},\ and\ \bibinfo {author} {\bibfnamefont {B.}~\bibnamefont {Mandelli}},\ }\bibfield  {title} {\bibinfo {title} {Studies on {RPC} detectors operated with environmentally friendly gas mixtures in {LHC}-like conditions},\ }\href@noop {} {\bibfield  {journal} {\bibinfo  {journal} {Nuclear Instruments and Methods in Physics Research Section A: Accelerators, Spectrometers, Detectors and Associated Equipment}\ }\textbf {\bibinfo {volume} {1049}},\ \bibinfo {pages} {168097} (\bibinfo {year} {2023})}\BibitemShut {NoStop}%
\bibitem [{\citenamefont {Koch}\ and\ \citenamefont {Franck}(2015)}]{koch2015high}%
  \BibitemOpen
  \bibfield  {author} {\bibinfo {author} {\bibfnamefont {M.}~\bibnamefont {Koch}}\ and\ \bibinfo {author} {\bibfnamefont {C.}~\bibnamefont {Franck}},\ }\bibfield  {title} {\bibinfo {title} {High voltage insulation properties of {HFO}1234ze},\ }\href@noop {} {\bibfield  {journal} {\bibinfo  {journal} {IEEE Transactions on Dielectrics and Electrical Insulation}\ }\textbf {\bibinfo {volume} {22}},\ \bibinfo {pages} {3260} (\bibinfo {year} {2015})}\BibitemShut {NoStop}%
\bibitem [{\citenamefont {Chachereau}\ and\ \citenamefont {Franck}(2017)}]{alise2017proceedings}%
  \BibitemOpen
  \bibfield  {author} {\bibinfo {author} {\bibfnamefont {A.}~\bibnamefont {Chachereau}}\ and\ \bibinfo {author} {\bibfnamefont {C.}~\bibnamefont {Franck}},\ }\bibfield  {title} {\bibinfo {title} {Characterization of {HFO}1234ze mixtures with {N}$_{2}$ and {CO}$_{2}$ for use as gaseous electrical insulation media},\ }in\ \href@noop {} {\emph {\bibinfo {booktitle} {Proceedings of the 20th International Symposium on High Voltage Engineering (ISH 2017)}}}\ (\bibinfo {organization} {Cigr{\'e}},\ \bibinfo {year} {2017})\BibitemShut {NoStop}%
\bibitem [{\citenamefont {Chachereau}\ \emph {et~al.}(2016)\citenamefont {Chachereau}, \citenamefont {Rabie},\ and\ \citenamefont {Franck}}]{alise2016paper}%
  \BibitemOpen
  \bibfield  {author} {\bibinfo {author} {\bibfnamefont {A.}~\bibnamefont {Chachereau}}, \bibinfo {author} {\bibfnamefont {M.}~\bibnamefont {Rabie}},\ and\ \bibinfo {author} {\bibfnamefont {C.~M.}\ \bibnamefont {Franck}},\ }\bibfield  {title} {\bibinfo {title} {Electron swarm parameters of the hydrofluoroolefine {HFO}1234ze},\ }\href {https://doi.org/10.1088/0963-0252/25/4/045005} {\bibfield  {journal} {\bibinfo  {journal} {Plasma Sources Science and Technology}\ }\textbf {\bibinfo {volume} {25}},\ \bibinfo {pages} {045005} (\bibinfo {year} {2016})}\BibitemShut {NoStop}%
\bibitem [{\citenamefont {Dahl}\ \emph {et~al.}(2012)\citenamefont {Dahl}, \citenamefont {Teich},\ and\ \citenamefont {Franck}}]{dahl2012obtaining}%
  \BibitemOpen
  \bibfield  {author} {\bibinfo {author} {\bibfnamefont {D.~A.}\ \bibnamefont {Dahl}}, \bibinfo {author} {\bibfnamefont {T.~H.}\ \bibnamefont {Teich}},\ and\ \bibinfo {author} {\bibfnamefont {C.~M.}\ \bibnamefont {Franck}},\ }\bibfield  {title} {\bibinfo {title} {Obtaining precise electron swarm parameters from a pulsed {T}ownsend setup},\ }\href@noop {} {\bibfield  {journal} {\bibinfo  {journal} {Journal of Physics D: Applied Physics}\ }\textbf {\bibinfo {volume} {45}},\ \bibinfo {pages} {485201} (\bibinfo {year} {2012})}\BibitemShut {NoStop}%
\bibitem [{\citenamefont {{\v{S}}a{\v{s}}i{\'c}}\ \emph {et~al.}(2013)\citenamefont {{\v{S}}a{\v{s}}i{\'c}}, \citenamefont {Dupljanin}, \citenamefont {de~Urquijo},\ and\ \citenamefont {Petrovi{\'c}}}]{sasic2013scattering}%
  \BibitemOpen
  \bibfield  {author} {\bibinfo {author} {\bibfnamefont {O.}~\bibnamefont {{\v{S}}a{\v{s}}i{\'c}}}, \bibinfo {author} {\bibfnamefont {S.}~\bibnamefont {Dupljanin}}, \bibinfo {author} {\bibfnamefont {J.}~\bibnamefont {de~Urquijo}},\ and\ \bibinfo {author} {\bibfnamefont {Z.~L.}\ \bibnamefont {Petrovi{\'c}}},\ }\bibfield  {title} {\bibinfo {title} {Scattering cross sections for electrons in {C}$_{2}${H}$_{2}${F}$_{4}$ and its mixtures with {A}r from measured transport coefficients},\ }\href@noop {} {\bibfield  {journal} {\bibinfo  {journal} {Journal of Physics D: Applied Physics}\ }\textbf {\bibinfo {volume} {46}},\ \bibinfo {pages} {325201} (\bibinfo {year} {2013})}\BibitemShut {NoStop}%
\bibitem [{\citenamefont {Zhang}\ \emph {et~al.}(2023)\citenamefont {Zhang}, \citenamefont {Hao}, \citenamefont {Yao}, \citenamefont {Xiong}, \citenamefont {Li}, \citenamefont {Murphy}, \citenamefont {Sinha}, \citenamefont {Antony},\ and\ \citenamefont {Ambalampitiya}}]{zhang2023determination}%
  \BibitemOpen
  \bibfield  {author} {\bibinfo {author} {\bibfnamefont {B.}~\bibnamefont {Zhang}}, \bibinfo {author} {\bibfnamefont {M.}~\bibnamefont {Hao}}, \bibinfo {author} {\bibfnamefont {Y.}~\bibnamefont {Yao}}, \bibinfo {author} {\bibfnamefont {J.}~\bibnamefont {Xiong}}, \bibinfo {author} {\bibfnamefont {X.}~\bibnamefont {Li}}, \bibinfo {author} {\bibfnamefont {A.~B.}\ \bibnamefont {Murphy}}, \bibinfo {author} {\bibfnamefont {N.}~\bibnamefont {Sinha}}, \bibinfo {author} {\bibfnamefont {B.}~\bibnamefont {Antony}},\ and\ \bibinfo {author} {\bibfnamefont {H.~B.}\ \bibnamefont {Ambalampitiya}},\ }\bibfield  {title} {\bibinfo {title} {Determination and assessment of a complete and self-consistent electron-neutral collision cross-section set for the {C}$_{4}${F}$_{7}${N} molecule},\ }\href@noop {} {\bibfield  {journal} {\bibinfo  {journal} {Journal of Physics D: Applied Physics}\ }\textbf {\bibinfo {volume} {56}},\ \bibinfo {pages} {134001} (\bibinfo {year} {2023})}\BibitemShut {NoStop}%
\bibitem [{\citenamefont {Kimura}\ and\ \citenamefont {Nakamura}(2010)}]{kimura2010electron}%
  \BibitemOpen
  \bibfield  {author} {\bibinfo {author} {\bibfnamefont {M.}~\bibnamefont {Kimura}}\ and\ \bibinfo {author} {\bibfnamefont {Y.}~\bibnamefont {Nakamura}},\ }\bibfield  {title} {\bibinfo {title} {Electron swarm parameters in {CF}$_{3}${I} and a set of electron collision cross sections for the {CF}$_{3}${I} molecule},\ }\href@noop {} {\bibfield  {journal} {\bibinfo  {journal} {Journal of Physics D: Applied Physics}\ }\textbf {\bibinfo {volume} {43}},\ \bibinfo {pages} {145202} (\bibinfo {year} {2010})}\BibitemShut {NoStop}%
\bibitem [{\citenamefont {Kumar}(1984)}]{kumar1984physics}%
  \BibitemOpen
  \bibfield  {author} {\bibinfo {author} {\bibfnamefont {K.}~\bibnamefont {Kumar}},\ }\bibfield  {title} {\bibinfo {title} {The physics of swarms and some basic questions of kinetic theory},\ }\href@noop {} {\bibfield  {journal} {\bibinfo  {journal} {Physics Reports}\ }\textbf {\bibinfo {volume} {112}},\ \bibinfo {pages} {319} (\bibinfo {year} {1984})}\BibitemShut {NoStop}%
\bibitem [{\citenamefont {Robson}\ and\ \citenamefont {Ness}(1986)}]{robson1986velocity}%
  \BibitemOpen
  \bibfield  {author} {\bibinfo {author} {\bibfnamefont {R.}~\bibnamefont {Robson}}\ and\ \bibinfo {author} {\bibfnamefont {K.}~\bibnamefont {Ness}},\ }\bibfield  {title} {\bibinfo {title} {Velocity distribution function and transport coefficients of electron swarms in gases: spherical-harmonics decomposition of {B}oltzmann’s equation},\ }\href@noop {} {\bibfield  {journal} {\bibinfo  {journal} {Physical Review A}\ }\textbf {\bibinfo {volume} {33}},\ \bibinfo {pages} {2068} (\bibinfo {year} {1986})}\BibitemShut {NoStop}%
\bibitem [{\citenamefont {Morgan}\ and\ \citenamefont {Penetrante}(1990)}]{morgan1990elendif}%
  \BibitemOpen
  \bibfield  {author} {\bibinfo {author} {\bibfnamefont {W.}~\bibnamefont {Morgan}}\ and\ \bibinfo {author} {\bibfnamefont {B.}~\bibnamefont {Penetrante}},\ }\bibfield  {title} {\bibinfo {title} {{ELENDIF}: A time-dependent {B}oltzmann solver for partially ionized plasmas},\ }\href@noop {} {\bibfield  {journal} {\bibinfo  {journal} {Computer Physics Communications}\ }\textbf {\bibinfo {volume} {58}},\ \bibinfo {pages} {127} (\bibinfo {year} {1990})}\BibitemShut {NoStop}%
\bibitem [{\citenamefont {Hagelaar}\ and\ \citenamefont {Pitchford}(2005)}]{hagelaar2005solving}%
  \BibitemOpen
  \bibfield  {author} {\bibinfo {author} {\bibfnamefont {G.}~\bibnamefont {Hagelaar}}\ and\ \bibinfo {author} {\bibfnamefont {L.}~\bibnamefont {Pitchford}},\ }\bibfield  {title} {\bibinfo {title} {Solving the {B}oltzmann equation to obtain electron transport coefficients and rate coefficients for fluid models},\ }\href@noop {} {\bibfield  {journal} {\bibinfo  {journal} {Plasma Sources Science and Technology}\ }\textbf {\bibinfo {volume} {14}},\ \bibinfo {pages} {722} (\bibinfo {year} {2005})}\BibitemShut {NoStop}%
\bibitem [{\citenamefont {Fraser}\ and\ \citenamefont {Mathieson}(1986)}]{fraser1986monte}%
  \BibitemOpen
  \bibfield  {author} {\bibinfo {author} {\bibfnamefont {G.}~\bibnamefont {Fraser}}\ and\ \bibinfo {author} {\bibfnamefont {E.}~\bibnamefont {Mathieson}},\ }\bibfield  {title} {\bibinfo {title} {{M}onte {C}arlo calculation of electron transport coefficients in counting gas mixtures: argon-methane mixtures},\ }\href@noop {} {\bibfield  {journal} {\bibinfo  {journal} {Nuclear Instruments and Methods in Physics Research Section A: Accelerators, Spectrometers, Detectors and Associated Equipment}\ }\textbf {\bibinfo {volume} {247}},\ \bibinfo {pages} {544} (\bibinfo {year} {1986})}\BibitemShut {NoStop}%
\bibitem [{\citenamefont {Biagi}(1999)}]{biagi1999monte}%
  \BibitemOpen
  \bibfield  {author} {\bibinfo {author} {\bibfnamefont {S.}~\bibnamefont {Biagi}},\ }\bibfield  {title} {\bibinfo {title} {Monte {C}arlo simulation of electron drift and diffusion in counting gases under the influence of electric and magnetic fields},\ }\href@noop {} {\bibfield  {journal} {\bibinfo  {journal} {Nuclear Instruments and Methods in Physics Research Section A: Accelerators, Spectrometers, Detectors and Associated Equipment}\ }\textbf {\bibinfo {volume} {421}},\ \bibinfo {pages} {234} (\bibinfo {year} {1999})}\BibitemShut {NoStop}%
\bibitem [{\citenamefont {Rabie}\ and\ \citenamefont {Franck}(2016)}]{rabie2016methes}%
  \BibitemOpen
  \bibfield  {author} {\bibinfo {author} {\bibfnamefont {M.}~\bibnamefont {Rabie}}\ and\ \bibinfo {author} {\bibfnamefont {C.~M.}\ \bibnamefont {Franck}},\ }\bibfield  {title} {\bibinfo {title} {{METHES}: A {M}onte {C}arlo collision code for the simulation of electron transport in low temperature plasmas},\ }\href@noop {} {\bibfield  {journal} {\bibinfo  {journal} {Computer Physics Communications}\ }\textbf {\bibinfo {volume} {203}},\ \bibinfo {pages} {268} (\bibinfo {year} {2016})}\BibitemShut {NoStop}%
\bibitem [{\citenamefont {Dupljanin}\ \emph {et~al.}(2010)\citenamefont {Dupljanin}, \citenamefont {de~Urquijo}, \citenamefont {{\v{S}}a{\v{s}}i{\'c}}, \citenamefont {Basurto}, \citenamefont {Ju{\'a}rez}, \citenamefont {Hern{\'a}ndez-{\'A}vila}, \citenamefont {Dujko},\ and\ \citenamefont {Petrovi{\'c}}}]{dupljanin2010transport}%
  \BibitemOpen
  \bibfield  {author} {\bibinfo {author} {\bibfnamefont {S.}~\bibnamefont {Dupljanin}}, \bibinfo {author} {\bibfnamefont {J.}~\bibnamefont {de~Urquijo}}, \bibinfo {author} {\bibfnamefont {O.}~\bibnamefont {{\v{S}}a{\v{s}}i{\'c}}}, \bibinfo {author} {\bibfnamefont {E.}~\bibnamefont {Basurto}}, \bibinfo {author} {\bibfnamefont {A.}~\bibnamefont {Ju{\'a}rez}}, \bibinfo {author} {\bibfnamefont {J.}~\bibnamefont {Hern{\'a}ndez-{\'A}vila}}, \bibinfo {author} {\bibfnamefont {S.}~\bibnamefont {Dujko}},\ and\ \bibinfo {author} {\bibfnamefont {Z.~L.}\ \bibnamefont {Petrovi{\'c}}},\ }\bibfield  {title} {\bibinfo {title} {Transport coefficients and cross sections for electrons in {N}$_{2}${O} and {N}$_{2}${O}/{N}$_{2}$ mixtures},\ }\href@noop {} {\bibfield  {journal} {\bibinfo  {journal} {Plasma Sources Science and Technology}\ }\textbf {\bibinfo {volume} {19}},\ \bibinfo {pages} {025005} (\bibinfo {year} {2010})}\BibitemShut {NoStop}%
\bibitem [{\citenamefont {Crompton}(1994)}]{crompton1994benchmark}%
  \BibitemOpen
  \bibfield  {author} {\bibinfo {author} {\bibfnamefont {R.}~\bibnamefont {Crompton}},\ }\bibfield  {title} {\bibinfo {title} {Benchmark measurements of cross sections for electron collisions: electron swarm methods},\ }\href@noop {} {\bibfield  {journal} {\bibinfo  {journal} {Advances in {A}tomic, {M}olecular, and {O}ptical {P}hysics}\ }\textbf {\bibinfo {volume} {33}},\ \bibinfo {pages} {97} (\bibinfo {year} {1994})}\BibitemShut {NoStop}%
\bibitem [{\citenamefont {Robson}\ \emph {et~al.}(1997)\citenamefont {Robson}, \citenamefont {Hildebrandt},\ and\ \citenamefont {Schmidt}}]{robson1997electron}%
  \BibitemOpen
  \bibfield  {author} {\bibinfo {author} {\bibfnamefont {R.}~\bibnamefont {Robson}}, \bibinfo {author} {\bibfnamefont {M.}~\bibnamefont {Hildebrandt}},\ and\ \bibinfo {author} {\bibfnamefont {B.}~\bibnamefont {Schmidt}},\ }\bibfield  {title} {\bibinfo {title} {Electron transport theory in gases: must it be so difficult?},\ }\href@noop {} {\bibfield  {journal} {\bibinfo  {journal} {Nuclear Instruments and Methods in Physics Research Section A: Accelerators, Spectrometers, Detectors and Associated Equipment}\ }\textbf {\bibinfo {volume} {394}},\ \bibinfo {pages} {74} (\bibinfo {year} {1997})}\BibitemShut {NoStop}%
\bibitem [{\citenamefont {Petrovi{\'c}}\ \emph {et~al.}(2009)\citenamefont {Petrovi{\'c}}, \citenamefont {Dujko}, \citenamefont {Mari{\'c}}, \citenamefont {Malovi{\'c}}, \citenamefont {Nikitovi{\'c}}, \citenamefont {{\v{S}}a{\v{s}}i{\'c}}, \citenamefont {Jovanovi{\'c}}, \citenamefont {Stojanovi{\'c}},\ and\ \citenamefont {Radmilovi{\'c}-Raenovi{\'c}}}]{petrovic2009measurement}%
  \BibitemOpen
  \bibfield  {author} {\bibinfo {author} {\bibfnamefont {Z.~L.}\ \bibnamefont {Petrovi{\'c}}}, \bibinfo {author} {\bibfnamefont {S.}~\bibnamefont {Dujko}}, \bibinfo {author} {\bibfnamefont {D.}~\bibnamefont {Mari{\'c}}}, \bibinfo {author} {\bibfnamefont {G.}~\bibnamefont {Malovi{\'c}}}, \bibinfo {author} {\bibfnamefont {{\v{Z}}.}~\bibnamefont {Nikitovi{\'c}}}, \bibinfo {author} {\bibfnamefont {O.}~\bibnamefont {{\v{S}}a{\v{s}}i{\'c}}}, \bibinfo {author} {\bibfnamefont {J.}~\bibnamefont {Jovanovi{\'c}}}, \bibinfo {author} {\bibfnamefont {V.}~\bibnamefont {Stojanovi{\'c}}},\ and\ \bibinfo {author} {\bibfnamefont {M.}~\bibnamefont {Radmilovi{\'c}-Raenovi{\'c}}},\ }\bibfield  {title} {\bibinfo {title} {Measurement and interpretation of swarm parameters and their application in plasma modelling},\ }\href@noop {} {\bibfield  {journal} {\bibinfo  {journal} {Journal of Physics D: Applied Physics}\ }\textbf {\bibinfo {volume} {42}},\ \bibinfo {pages} {194002} (\bibinfo {year} {2009})}\BibitemShut {NoStop}%
\bibitem [{\citenamefont {Sasic}\ \emph {et~al.}(2010)\citenamefont {Sasic}, \citenamefont {de~Urquijo}, \citenamefont {Ju{\'a}rez}, \citenamefont {Dupljanin}, \citenamefont {Jovanovic}, \citenamefont {Hern{\'a}ndez-{\'A}vila}, \citenamefont {Basurto},\ and\ \citenamefont {Petrovic}}]{sasic2010measurements}%
  \BibitemOpen
  \bibfield  {author} {\bibinfo {author} {\bibfnamefont {O.}~\bibnamefont {Sasic}}, \bibinfo {author} {\bibfnamefont {J.}~\bibnamefont {de~Urquijo}}, \bibinfo {author} {\bibfnamefont {A.}~\bibnamefont {Ju{\'a}rez}}, \bibinfo {author} {\bibfnamefont {S.}~\bibnamefont {Dupljanin}}, \bibinfo {author} {\bibfnamefont {J.}~\bibnamefont {Jovanovic}}, \bibinfo {author} {\bibfnamefont {J.}~\bibnamefont {Hern{\'a}ndez-{\'A}vila}}, \bibinfo {author} {\bibfnamefont {E.}~\bibnamefont {Basurto}},\ and\ \bibinfo {author} {\bibfnamefont {Z.~L.}\ \bibnamefont {Petrovic}},\ }\bibfield  {title} {\bibinfo {title} {Measurements and analysis of electron transport coefficients obtained by a pulsed {T}ownsend technique},\ }\href@noop {} {\bibfield  {journal} {\bibinfo  {journal} {Plasma Sources Science Technology}\ }\textbf {\bibinfo {volume} {19}} (\bibinfo {year} {2010})}\BibitemShut {NoStop}%
\bibitem [{\citenamefont {Benussi}\ \emph {et~al.}(2015)\citenamefont {Benussi}, \citenamefont {Bianco}, \citenamefont {Piccolo}, \citenamefont {Saviano}, \citenamefont {Colafranceschi}, \citenamefont {Kj{\o}lbro}, \citenamefont {Sharma}, \citenamefont {Yang}, \citenamefont {Chen}, \citenamefont {Ban} \emph {et~al.}}]{benussi2015properties}%
  \BibitemOpen
  \bibfield  {author} {\bibinfo {author} {\bibfnamefont {L.}~\bibnamefont {Benussi}}, \bibinfo {author} {\bibfnamefont {S.}~\bibnamefont {Bianco}}, \bibinfo {author} {\bibfnamefont {D.}~\bibnamefont {Piccolo}}, \bibinfo {author} {\bibfnamefont {G.}~\bibnamefont {Saviano}}, \bibinfo {author} {\bibfnamefont {S.}~\bibnamefont {Colafranceschi}}, \bibinfo {author} {\bibfnamefont {J.}~\bibnamefont {Kj{\o}lbro}}, \bibinfo {author} {\bibfnamefont {A.}~\bibnamefont {Sharma}}, \bibinfo {author} {\bibfnamefont {D.}~\bibnamefont {Yang}}, \bibinfo {author} {\bibfnamefont {G.}~\bibnamefont {Chen}}, \bibinfo {author} {\bibfnamefont {Y.}~\bibnamefont {Ban}}, \emph {et~al.},\ }\bibfield  {title} {\bibinfo {title} {Properties of potential eco-friendly gas replacements for particle detectors in high-energy physics},\ }\href@noop {} {\bibfield  {journal} {\bibinfo  {journal} {arXiv preprint arXiv:1505.00701}\ } (\bibinfo {year} {2015})}\BibitemShut {NoStop}%
\bibitem [{\citenamefont {Christophorou}\ and\ \citenamefont {Hunter}(1984)}]{christophorou1984electrons}%
  \BibitemOpen
  \bibfield  {author} {\bibinfo {author} {\bibfnamefont {L.}~\bibnamefont {Christophorou}}\ and\ \bibinfo {author} {\bibfnamefont {S.}~\bibnamefont {Hunter}},\ }\bibfield  {title} {\bibinfo {title} {Electrons in dense gases},\ }in\ \href@noop {} {\emph {\bibinfo {booktitle} {Swarms of Ions and Electrons in Gases}}}\ (\bibinfo  {publisher} {Springer},\ \bibinfo {year} {1984})\ pp.\ \bibinfo {pages} {241--264}\BibitemShut {NoStop}%
\bibitem [{\citenamefont {Bloch}\ and\ \citenamefont {Bradbury}(1935)}]{bloch1935mechanism}%
  \BibitemOpen
  \bibfield  {author} {\bibinfo {author} {\bibfnamefont {F.}~\bibnamefont {Bloch}}\ and\ \bibinfo {author} {\bibfnamefont {N.~E.}\ \bibnamefont {Bradbury}},\ }\bibfield  {title} {\bibinfo {title} {On the mechanism of unimolecular electron capture},\ }\href@noop {} {\bibfield  {journal} {\bibinfo  {journal} {Physical Review}\ }\textbf {\bibinfo {volume} {48}},\ \bibinfo {pages} {689} (\bibinfo {year} {1935})}\BibitemShut {NoStop}%
\bibitem [{\citenamefont {Aleksandrov}(1988)}]{aleksandrov1988three}%
  \BibitemOpen
  \bibfield  {author} {\bibinfo {author} {\bibfnamefont {N.}~\bibnamefont {Aleksandrov}},\ }\bibfield  {title} {\bibinfo {title} {Three-body electron attachment to a molecule},\ }\href@noop {} {\bibfield  {journal} {\bibinfo  {journal} {Soviet Physics Uspekhi}\ }\textbf {\bibinfo {volume} {31}},\ \bibinfo {pages} {101} (\bibinfo {year} {1988})}\BibitemShut {NoStop}%
\bibitem [{\citenamefont {White}\ \emph {et~al.}(2003)\citenamefont {White}, \citenamefont {Robson}, \citenamefont {Schmidt},\ and\ \citenamefont {Morrison}}]{white2003classical}%
  \BibitemOpen
  \bibfield  {author} {\bibinfo {author} {\bibfnamefont {R.~D.}\ \bibnamefont {White}}, \bibinfo {author} {\bibfnamefont {R.~E.}\ \bibnamefont {Robson}}, \bibinfo {author} {\bibfnamefont {B.}~\bibnamefont {Schmidt}},\ and\ \bibinfo {author} {\bibfnamefont {M.~A.}\ \bibnamefont {Morrison}},\ }\bibfield  {title} {\bibinfo {title} {Is the classical two-term approximation of electron kinetic theory satisfactory for swarms and plasmas?},\ }\href@noop {} {\bibfield  {journal} {\bibinfo  {journal} {Journal of Physics D: Applied Physics}\ }\textbf {\bibinfo {volume} {36}},\ \bibinfo {pages} {3125} (\bibinfo {year} {2003})}\BibitemShut {NoStop}%
\bibitem [{\citenamefont {Pancheshnyi}\ \emph {et~al.}(2012)\citenamefont {Pancheshnyi}, \citenamefont {Biagi}, \citenamefont {Bordage}, \citenamefont {Hagelaar}, \citenamefont {Morgan}, \citenamefont {Phelps},\ and\ \citenamefont {Pitchford}}]{projectlxcat}%
  \BibitemOpen
  \bibfield  {author} {\bibinfo {author} {\bibfnamefont {S.}~\bibnamefont {Pancheshnyi}}, \bibinfo {author} {\bibfnamefont {S.}~\bibnamefont {Biagi}}, \bibinfo {author} {\bibfnamefont {M.}~\bibnamefont {Bordage}}, \bibinfo {author} {\bibfnamefont {G.}~\bibnamefont {Hagelaar}}, \bibinfo {author} {\bibfnamefont {W.}~\bibnamefont {Morgan}}, \bibinfo {author} {\bibfnamefont {A.}~\bibnamefont {Phelps}},\ and\ \bibinfo {author} {\bibfnamefont {L.}~\bibnamefont {Pitchford}},\ }\bibfield  {title} {\bibinfo {title} {The {LXC}at project: Electron scattering cross sections and swarm parameters for low temperature plasma modeling},\ }\href@noop {} {\bibfield  {journal} {\bibinfo  {journal} {Chemical Physics}\ }\textbf {\bibinfo {volume} {398}},\ \bibinfo {pages} {148} (\bibinfo {year} {2012})}\BibitemShut {NoStop}%
\bibitem [{\citenamefont {Bianchi}\ \emph {et~al.}(2020{\natexlab{a}})\citenamefont {Bianchi}, \citenamefont {Delsanto}, \citenamefont {Dupieux}, \citenamefont {Ferretti}, \citenamefont {Gagliardi}, \citenamefont {Joly}, \citenamefont {Manen}, \citenamefont {Marchisone}, \citenamefont {Micheletti}, \citenamefont {Rosano} \emph {et~al.}}]{bianchi2020siena}%
  \BibitemOpen
  \bibfield  {author} {\bibinfo {author} {\bibfnamefont {A.}~\bibnamefont {Bianchi}}, \bibinfo {author} {\bibfnamefont {S.}~\bibnamefont {Delsanto}}, \bibinfo {author} {\bibfnamefont {P.}~\bibnamefont {Dupieux}}, \bibinfo {author} {\bibfnamefont {A.}~\bibnamefont {Ferretti}}, \bibinfo {author} {\bibfnamefont {M.}~\bibnamefont {Gagliardi}}, \bibinfo {author} {\bibfnamefont {B.}~\bibnamefont {Joly}}, \bibinfo {author} {\bibfnamefont {S.}~\bibnamefont {Manen}}, \bibinfo {author} {\bibfnamefont {M.}~\bibnamefont {Marchisone}}, \bibinfo {author} {\bibfnamefont {L.}~\bibnamefont {Micheletti}}, \bibinfo {author} {\bibfnamefont {A.}~\bibnamefont {Rosano}}, \emph {et~al.},\ }\bibfield  {title} {\bibinfo {title} {Studies on tetrafluoropropene-based gas mixtures with low environmental impact for {R}esistive {P}late {C}hambers},\ }\href@noop {} {\bibfield  {journal} {\bibinfo  {journal} {Journal of Instrumentation}\ }\textbf {\bibinfo {volume} {15}}\bibinfo  {number} { (04)},\ \bibinfo {pages} {C04039}}\BibitemShut
  {NoStop}%
\bibitem [{\citenamefont {Bianchi}(2023)}]{bianchi2023matoq}%
  \BibitemOpen
\bibfield  {number} {  }\bibfield  {author} {\bibinfo {author} {\bibfnamefont {A.}~\bibnamefont {Bianchi}},\ }\bibfield  {title} {\bibinfo {title} {{MATOQ}: a {M}onte {C}arlo simulation of electron transport in environmental-friendly gas mixtures for {R}esistive {P}late {C}hambers},\ }\bibfield  {journal} {\bibinfo  {journal} {The European Physical Journal Plus}\ }\textbf {\bibinfo {volume} {138}},\ \href {https://doi.org/10.1140/epjp/s13360-023-04440-0} {10.1140/epjp/s13360-023-04440-0} (\bibinfo {year} {2023})\BibitemShut {NoStop}%
\bibitem [{\citenamefont {Bianchi}\ \emph {et~al.}(2020{\natexlab{b}})\citenamefont {Bianchi}, \citenamefont {Delsanto}, \citenamefont {Dupieux}, \citenamefont {Ferretti}, \citenamefont {Gagliardi}, \citenamefont {Joly}, \citenamefont {Manen}, \citenamefont {Marchisone}, \citenamefont {Micheletti}, \citenamefont {Quaglia} \emph {et~al.}}]{bianchiRPC2020}%
  \BibitemOpen
  \bibfield  {author} {\bibinfo {author} {\bibfnamefont {A.}~\bibnamefont {Bianchi}}, \bibinfo {author} {\bibfnamefont {S.}~\bibnamefont {Delsanto}}, \bibinfo {author} {\bibfnamefont {P.}~\bibnamefont {Dupieux}}, \bibinfo {author} {\bibfnamefont {A.}~\bibnamefont {Ferretti}}, \bibinfo {author} {\bibfnamefont {M.}~\bibnamefont {Gagliardi}}, \bibinfo {author} {\bibfnamefont {B.}~\bibnamefont {Joly}}, \bibinfo {author} {\bibfnamefont {S.}~\bibnamefont {Manen}}, \bibinfo {author} {\bibfnamefont {M.}~\bibnamefont {Marchisone}}, \bibinfo {author} {\bibfnamefont {L.}~\bibnamefont {Micheletti}}, \bibinfo {author} {\bibfnamefont {L.}~\bibnamefont {Quaglia}}, \emph {et~al.},\ }\bibfield  {title} {\bibinfo {title} {Environment-friendly gas mixtures for {R}esistive {P}late {C}hambers: an experimental and simulation study},\ }\href@noop {} {\bibfield  {journal} {\bibinfo  {journal} {Journal of Instrumentation}\ }\textbf {\bibinfo {volume} {15}}\bibinfo  {number} { (09)},\ \bibinfo {pages} {C09006}}\BibitemShut {NoStop}%
\end{thebibliography}%

\end{document}